\documentclass[preprint,12pt,authoryear]{elsarticle}

\usepackage{amssymb}
\usepackage{array,multirow}
 \usepackage{float}
 \usepackage{graphicx}
\usepackage{amsthm}
\usepackage{amsmath}
\usepackage{longtable}
\usepackage{lipsum}
\usepackage{gensymb}
\usepackage{blindtext}
\usepackage{caption}

\journal{Icarus}

\begin{document}

\begin{frontmatter}



\title{The Effect of Adsorbed Liquid and Material Density on Saltation Threshold: Insight from Laboratory and Wind Tunnel Experiments}

\author[1]{Xinting Yu\corref{cor1}}
\ead{xyu33@jhu.edu}
\author[1]{Sarah M. H\"orst}
\author[1]{Chao He}
\author[2]{Nathan T. Bridges}
\author[3]{Devon M. Burr}
\author[4]{Joshua A. Sebree}
\author[5]{James K. Smith}

\cortext[cor1]{Corresponding author at: Department of Earth and Planetary Sciences, Johns Hopkins University, Baltimore, Maryland 21218, USA}
\address[1]{Department of Earth and Planetary Sciences, Johns Hopkins University (xyu33@jhu.edu), Baltimore, Maryland 21218, USA}
\address[2]{Applied Physics Laboratory, Johns Hopkins University, Laurel, Maryland 20723, USA}
\address[3]{Department of Earth and Planetary Sciences, University of Tennessee-Knoxville, 306 EPS Building, 1412 Circle Drive, Knoxville, Tennessee 37996, USA}
\address[4]{Department of Chemistry and Biochemistry, University of Northern Iowa, Cedar Falls, Iowa 50614, USA}
\address[5]{Arizona State University, Tempe, AZ 85287-1404, USA}


\begin{abstract}
Saltation threshold, the minimum wind speed for sediment transport, is a fundamental parameter in aeolian processes. Measuring this threshold using boundary layer wind tunnels, in which particles are mobilized by flowing air, for a subset of different planetary conditions can inform our understanding of physical processes of sediment transport. The presence of liquid, such as water on Earth or methane on Titan, may affect the threshold values to a great extent. Sediment density is also crucial for determining threshold values. Here we provide quantitative data on density and water content of common wind tunnel materials (including chromite, basalt, quartz sand, beach sand, glass beads, gas chromatograph packing materials, walnut shells, iced tea powder, activated charcoal, instant coffee, and glass bubbles) that have been used to study conditions on Earth, Titan, Mars, and Venus. The measured density values for low density materials are higher compared to literature values (e.g., $\sim$30\% for walnut shells), whereas for the high density materials, there is no such discrepancy. We also find that low density materials have much higher water content and longer atmospheric equilibration timescales compared to high density sediments. We used thermogravimetric analysis (TGA) to quantify surface and internal water and found that over 80\% of the total water content is surface water for low density materials. In the Titan Wind Tunnel (TWT), where Reynolds number conditions similar to those on Titan can be achieved, we performed threshold experiments with the standard walnut shells (125--150 $\mathrm{\mu m}$, 7.2\% water by mass) and dried walnut shells, in which the water content was reduced to 1.7\%. The threshold results for the two scenarios are almost the same, which indicates that humidity had a negligible effect on threshold for walnut shells in this experimental regime. When the water content is lower than 11.0\%, the interparticle forces are dominated by adsorption forces, whereas at higher values the interparticle forces are dominated by much larger capillary forces. For materials with low equilibrium water content, like quartz sand, capillary forces dominate. When the interparticle forces are dominated by adsorption forces, the threshold does not increase with increasing relative humidity (RH) or water content. Only when the interparticle forces are dominated by capillary forces does the threshold start to increase with increasing RH/water content. Since tholins have a low methane content (0.3\% at saturation, Curtis et al., 2008), we believe tholins would behave similarly to quartz sand when subjected to methane moisture. 
\end{abstract}


\end{frontmatter}


\section{Introduction}
Aeolian processes are fundamental in modifying the surfaces of all solid bodies in the Solar System with permanent or ephemeral atmospheres, including Earth, Venus, Mars, Saturn's moon Titan (Greeley and  Iversen, 1985), Neptune's moon Triton (Smith et al., 1989), Pluto (Stern et al., 2015), and the comet 67P/Churyumov-Gerasimenko (Thomas et al., 2015). Studying aeolian features on planetary bodies enhances our understanding of near-surface winds, including the minimum wind speed to initiate saltation, wind direction, sediment flux, dune migration rates, and landscape modification. This information also provides input data and tests for global circulation predictions, leading to more powerful and accurate models. These models can then be run for different conditions providing insight into past or future climates.

Threshold wind speed is a fundamental parameter for understanding how and under what conditions wind detaches particles from the surface. Boundary layer wind tunnels serve as powerful laboratories for the study of aeolian processes, including threshold wind speed. Boundary layer tunnels were first used by Bagnold (1941), who pioneered the study of the minimum wind speed needed to initiate saltation on Earth. To test whether the parameters for quantifying threshold wind speed on Earth were also appropriate for Venusian and Martian conditions, the Martian Surface Wind Tunnel (MARSWIT) and Venus Wind Tunnel (VWT, now refurbished to the Titan Wind Tunnel, TWT) were built. The MARSWIT simulates the atmospheric pressure on Mars (4.0--8.7 mb) with both martian atmosphere ($\mathrm{CO_2}$) and dry air (Greeley et al., 1976, 1977, 1980). To simulate the weight of the grains under the lower gravity of Mars, low density materials like walnut shells have been used (Greeley et al., 1976). The VWT achieved the same atmospheric density as on Venus using $\mathrm{CO_2}$, and employed quartz sand as sediment (Greeley et al., 1984a, b).

Cassini spacecraft data show extensive linear dunes covering 35\% of equatorial regions ($\pm$30$\degree$) of Titan (Lorenz et al., 2006; Radebaugh et al., 2008). The dune materials are likely dominated by radar-dark (wavelength 2.17 cm) organic materials deposited from the atmosphere, with some minor water ice (McCord et al., 2006; Soderblom et al., 2007; Barnes et al., 2008; Clark et al., 2010; Le Gall et al., 2011; Hirtzig et al., 2013; Rodriguez et al., 2014). Global circulation models and measurements from the Huygens Doppler Wind Experiment show the dominant surface transporting winds are weak east to west winds (see e.g., Bird et al., 2005; Tokano, 2010). Conversely, the streamlined appearance of the dunes is consistent with west to east winds (see e.g., Lorenz et al., 2006). In order to address this mystery, as well as to quantify the threshold wind speed and to study other aeolian processes on Titan, the TWT was built (Burr et al., 2015a,b).

The robustness of wind tunnel experiments depends both on the degree of control of environmental conditions (e.g., pressure, relative humidity) and an understanding of experimental materials. The TWT simulates certain properties of Titan's near-surface atmosphere by using high pressure air (12.5 bar) to achieve the same Reynolds number (Re*) as on Titan. The Reynolds number is the ratio of inertial to viscous forces (Re*=u*$\mathrm{D_p/\nu}$, where u* is the threshold friction wind speed, $\mathrm{D_p}$ is particle size, and $\mathrm{\nu}$ is kinematic viscosity); this dimensionless number characterizes whether flow is laminar ($Re^*\ll1$) or turbulent ($Re^*\gg1$). Titan has a surface temperature of 94 K, a surface pressure of 1.5 bar (Lindal et al., 1983, Fulchignoni et al., 2005), and an estimated atmospheric kinematic viscosity that is only about 1/12th of Earth (Burr et al., 2015a, Extended Data Table 1). The TWT at 12.5 bars and room temperature achieves the same value for kinematic viscosity (6.25$\mathrm{\times10^{-6}\ Pa\cdot s}$) as on the surface of Titan. Low density materials have been used in threshold experiments in the TWT to compensate for Titan's low gravity, which is about $\mathrm{1/7^{th}}$ that of Earth ($\sim$ 1.4 m/$\mathrm{s^2}$). In the case of TWT experiments, a range of particle sizes and densities have been utilized to measure a wide span of potential conditions over which threshold can occur, thereby allowing extrapolation to the very low weight materials on Titan (Burr et al., 2015a).

The threshold wind speed is a function of the force balance of gravity ($\mathrm{F_g}$), aerodynamic lift ($\mathrm{F_l}$), aerodynamic drag ($\mathrm{F_d}$), and interparticle forces ($\mathrm{F_i}$), as shown in Fig. 1. $\mathrm{F_g}$ is dependent on mass which, in turn, is proportional to material density. For relatively heavy materials like quartz, the density determination is straight forward. However, for low density materials, most of which are porous and irregular, density values can vary considerably depending on the density definition used (see Section 3.1). The density values used in literature for low density materials (Greeley et al., 1976, 1977, 1980, Burr et al., 2015a) are usually taken from manufacturer labels, which usually do not specify a specific density definition. Thus it is necessary to reevaluate the data from these manufacturer labels.

Interparticle forces consist of van der Waals, cohesion, and electrostatic forces. Van der Waals forces describe the dipole-dipole interactions between neutral molecules. Cohesion forces are the attraction forces between particles with condensed liquid on them. Electrostatic forces are the attraction or repulsion forces between charged particles or particles with different surface potential. On Earth, where water is abundant, cohesion forces are much larger than van der Waals and electrostatic forces (McKenna Neuman et al., 2003). On Titan, electrostatic forces likely dominate the interparticle forces (see discussion in Lorenz 2014, Burr et al., 2015a), and the same may be true on other planetary bodies where liquid water is not abundant, such as Mars, Venus, Triton, Pluto, and Comet 67P. On Titan, cohesion between liquid ethane or methane could also be important (Lorenz 2014). The surface tensions ($\gamma_s$) of liquid methane and ethane at Titan's surface temperature and pressure are only about 15--20 mN/m (Baidakov et al. 2013) and are lower than water at 25$\degree$C on Earth ($\gamma_s$=72 mN/m). Thus the cohesion force for liquid methane and ethane on Titan should be lower than the cohesion for water on Earth given the same relative humidity for their respective surface temperatures.

Using the TWT, Burr et al. (2015a) found that the experimental saltation threshold wind speeds are 50\% higher than the model predictions of Iversen and White (1982) and Shao and  Lu (2000). The inclusion in these models of a density ratio term (Iversen et al., 1976) with the very low density ratio (sediment density over atmospheric density)  for Titan conditions caused the models to fit the TWT experimental data. Ongoing work includes using higher pressures in the TWT (15 and 20 bar) to further study the effect of the density ratio on threshold (Nield et al., 2016). Lower pressures are also used (1, 3, and 8 bar) to simulate possible past Titan conditions (possible past pressure as low as 0.7 bar, Charnay et al., 2014, Bridges et al., 2015).

The models used in the previous TWT work and many other wind tunnel experiments did not specify interparticle forces for different materials. The experiments also did not dry the materials, which were exposed to ambient atmosphere with relative humidity (RH) ranging from 45 to 65\%. During the TWT runs, the air used in the wind tunnel had an RH of 20 to 35\% in the current experimental regime. Thus the interparticle forces were likely dominated by cohesion forces because the electrostatic charges dissipate very quickly when RH is greater than 5\% (Bunker et al., 2007). In order to correctly simulate aeolian processes on Titan, where electrostatic forces are predicted to dominate the interparticle forces (Lorenz 2014, Burr et al., 2015a), a quantitative understanding of interparticle forces is therefore necessary. To accurately translate the TWT results to Titan conditions where liquid water is absent, we need to assess the effect of water present in Earth-based experiments on interparticle forces. Because of their low densities, which are used to provide some compensation for low extraterrestrial gravitational accelerations, the materials commonly used as analog sediments in Martian or Titan wind tunnel simulations are particularly susceptible to interparticle forces, highlighting the importance of this issue for understanding planetary aeolian processes.

\begin{figure}
\begin{center}
\includegraphics[width=\textwidth]{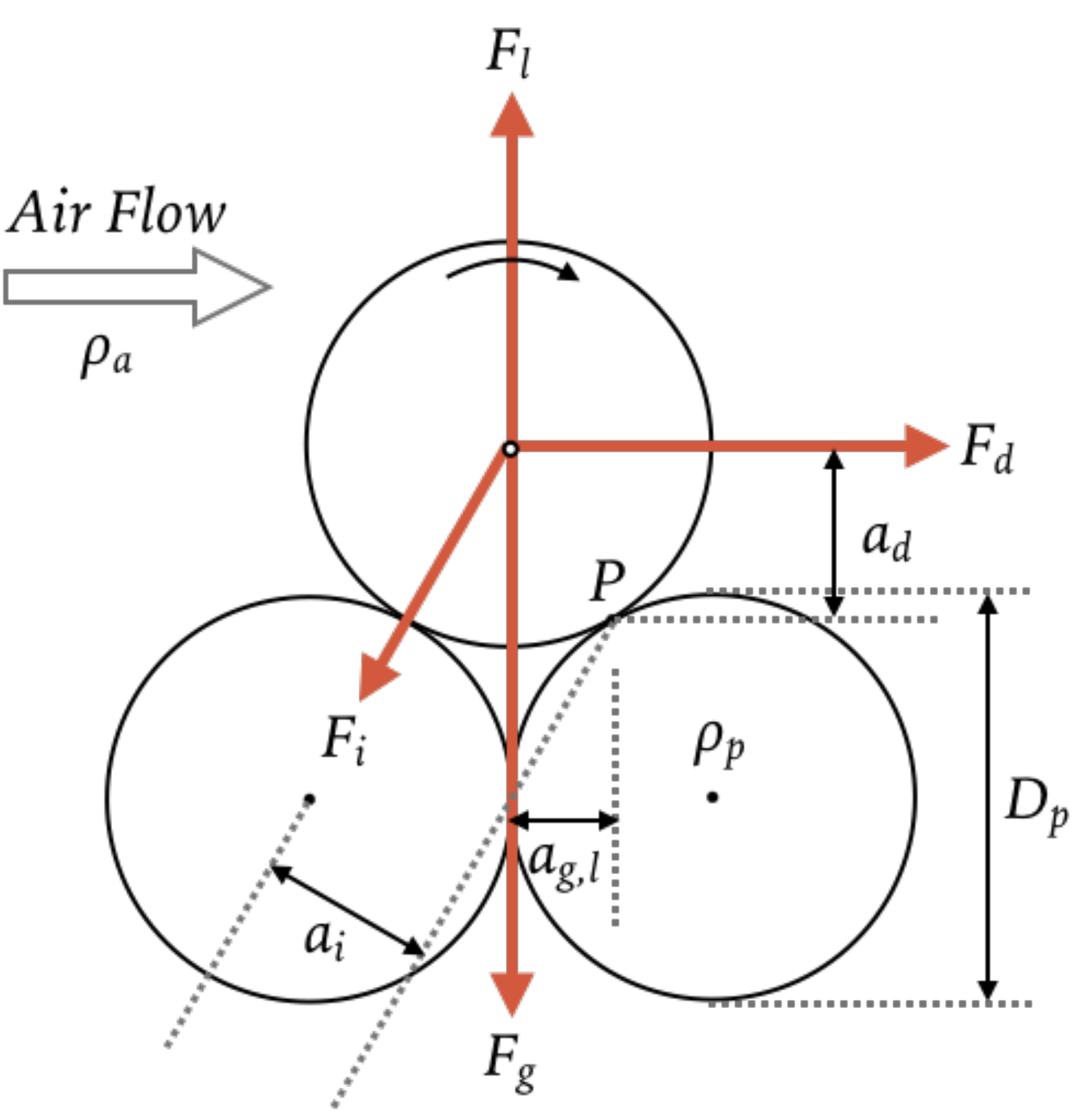}
\caption[]{Forces acting on a particle stacked on two particles in airstream of density $\mathrm{\rho_a}$ (after Shao and  Lu, 2000 and Kok et al., 2012). The particle density, $\mathrm{\rho_p}$, and the particle size, $\mathrm{D_p}$ are the same for all three particles. The forces include gravity ($\mathrm{F_g}$), the aerodynamic lift ($\mathrm{F_l}$), the aerodynamic drag ($\mathrm{F_d}$), and interparticle forces ($\mathrm{F_i}$). The moment arm lengths $\mathrm{a_g}$, $\mathrm{a_l}$, $\mathrm{a_d}$, and $\mathrm{a_i}$ correspond to $\mathrm{F_g}$, $\mathrm{F_l}$, $\mathrm{F_d}$, and $\mathrm{F_i}$, respectively.}
\label{fig:forcebalance}
\end{center}
\end{figure}

Previous studies of the effect of relative humidity on threshold which focused on Earth are reviewed below (Section 2). In Section 3.1, we summarize the common materials used in planetary wind tunnels and their basic properties according to the literature. The experimental methods are introduced in Section 3.2--3.4. We measured the density of materials in use in planetary wind tunnels (Section 4.1) and their gravimetric water content and Earth atmospheric equilibration timescales (Section 4.2). To further understand the effect of liquid on threshold, we measured the surface water content of the materials (Section 4.3). In Section 4.4, the threshold results of TWT experiments for wet and dry low density materials are shown. The implications for the threshold wind speed and entrainment of particles on Titan are discussed in Section 5.

\begin{table}[h!]
\centering
\caption{Summary of variables.}
\vspace{0.2cm}
 \begin{tabular}{|c|c|c|} 
 \hline
Variable Symbols & Description & Unit \\  
\hline\hline
u* & threshold friction wind speed & m/s\\
\hline
$\mathrm{D_p}$ & particle size & m\\
\hline
$\mathrm{\rho_a}$ & air density & kg/$\mathrm{m^3}$\\
\hline
$\mathrm{\rho_p}$ & particle density & kg/$\mathrm{m^3}$\\
\hline
$\mathrm{\mu}$ & dynamic viscosity & kg/(m$\cdot$s)\\
\hline
$\mathrm{\nu}$ & kinematic viscosity, $\mathrm{\mu/\rho_a}$ & $\mathrm{m^2}$/s\\
\hline
Re* & Reynolds number, u*$\mathrm{D_p/\nu}$ & -\\
\hline
g & gravity & m/$\mathrm{s^2}$\\ 
\hline
RH & relative humidity & \% by pressure\\ 
\hline
w & water content & \% by mass\\ 
\hline
w' & initiation water content & \% by mass\\ 
   \hline
\end{tabular}
\vspace{-0.2cm}
\label{table:variables}
\end{table}

\section{Previous studies of the effect of water on threshold}

Bagnold (1941) used the balance of gravity and aerodynamic drag to derive the threshold friction wind speed for dry sand:
\begin{equation}
\label{eq:bagnold}
u^*_b=A\sqrt{\frac{\rho_p-\rho_a}{\rho_a}gd}
\end{equation}
where A is a function of Reynolds number Re* and interparticle forces. When Re*\textgreater3.5 (when particles are beyond the viscous sublayer and are more susceptible to fluid drag), A is found to be a constant, with A=0.1 in air and A=0.2 in water, $\mathrm{\rho_p}$ and $\mathrm{\rho_a}$ are the density of the particle and atmosphere, respectively, and d is the mean aerodynamic particle diameter. This function is only appropriate for dry sand particles over 200 $\mathrm{\mu}$m; for smaller sediments, interparticle forces become more significant compared to the weight of the particles.

\begin{figure}
\begin{center}
\includegraphics[width=\textwidth]{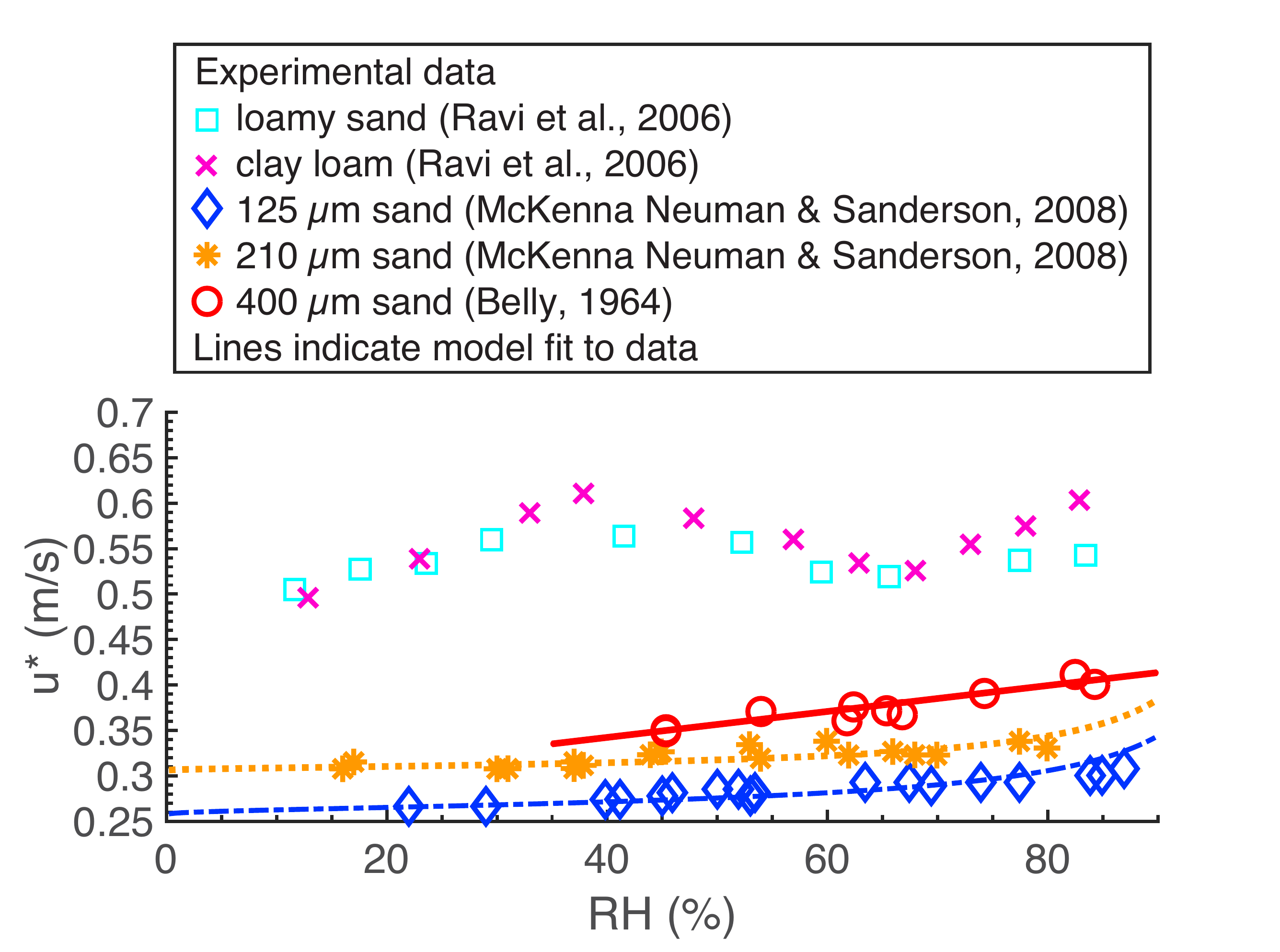}
\caption[]{Threshold friction wind speed (u*) variation with relative humidity (RH). The symbols represent experimental data for: (1) 400 $\mathrm{\mu}$m sand (Belly, 1964); (2) 125 $\mathrm{\mu}$m sand (McKenna Neuman and Sanderson, 2008); (3) 210 $\mathrm{\mu}$m sand (McKenna Neuman and Sanderson, 2008); (4) loamy sand with 8\% soil clay content (Ravi et al., 2006); (5) clay loam with 31\% soil clay content (Ravi et al., 2006). The lines show model fits to the data: (1) 400 $\mathrm{\mu}$m sand using Equation (2); (2) 125 $\mathrm{\mu}$m sand using Equation (4); (3) 210 $\mathrm{\mu}$m sand using Equation (4).}
\label{fig:rhu}
\end{center}
\end{figure}

\begin{figure}
\begin{center}
\includegraphics[width=\textwidth]{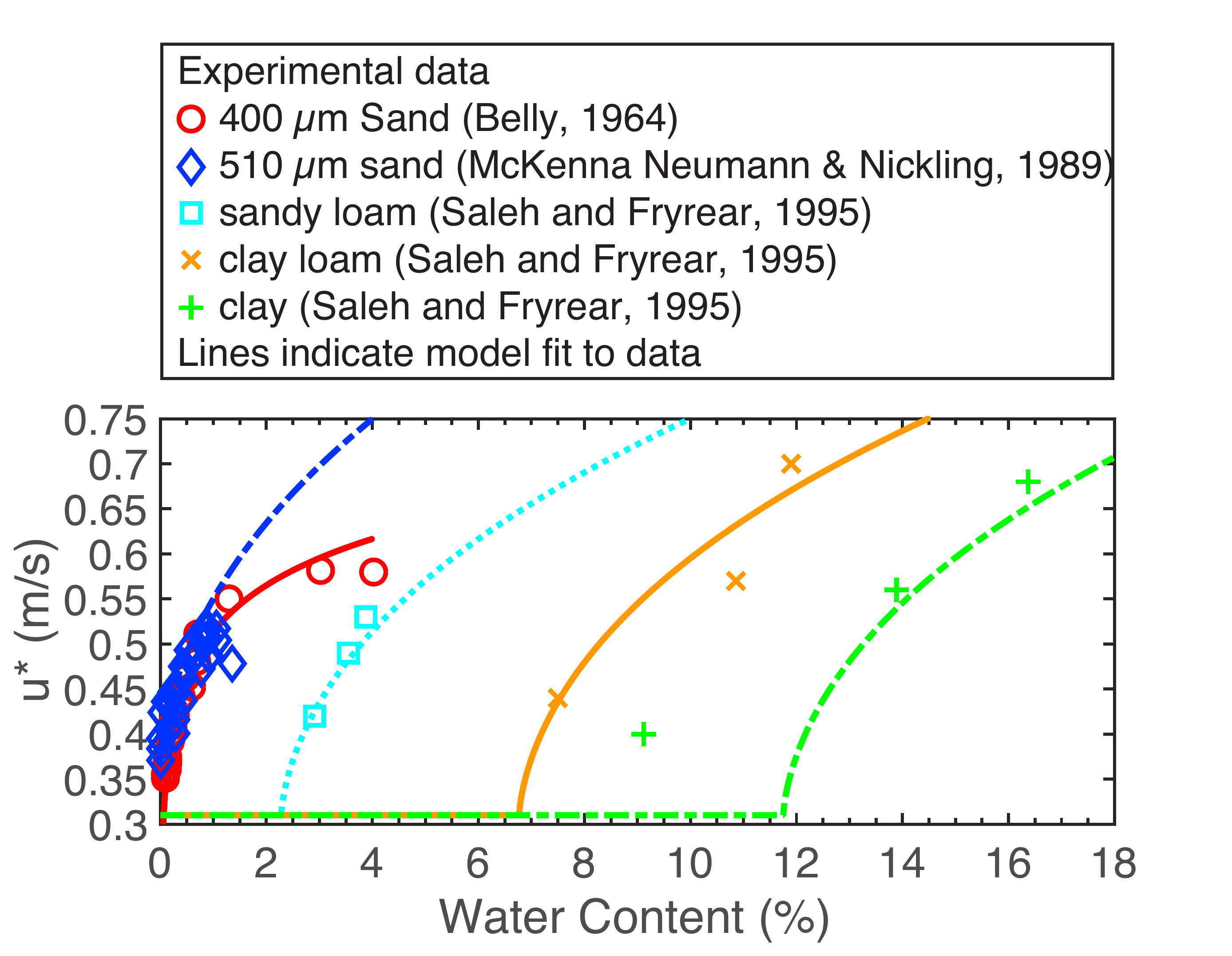}
\caption[]{Shown here is the threshold friction wind speed (u*) variation with water content. The symbols represent experimental data for: (1) 400 $\mathrm{\mu}$m sand (Belly, 1964); (2) 510 $\mathrm{\mu}$m sand (McKenna Neuman and Nickling, 1989); (3) sandy loam with 12.2 \% soil clay content (Saleh and Fryrear, 1995); (4) clay loam with 31.7 \% soil clay content (Saleh and Fryrear, 1995); (5) clay with 49.2\% soil clay content (Saleh and Fryrear, 1995). The lines show model fit to the data: (1) 400 $\mathrm{\mu}$m sand using Equation (2); (2) 510 $\mathrm{\mu}$m sand using Equation (10); (3) sandy loam using Equation (10); (4) clay loam using Equation (10); (5) clay using Equation (10).}
\label{fig:watercontentu}
\end{center}
\end{figure}

Belly (1964) conducted the first wind tunnel experiments on the effect of humidity on threshold using 400 $\mu$m sand, and found that,
\begin{equation}
\label{eq:belly}
u^*_w=u^*_b(1.8+0.6\ \mathrm{log}\ w)=u^*_b(1+\frac{1}{2}\frac{\mathrm{RH}}{100}),
\end{equation}
where $\mathrm{u^*_w}$ stands for threshold for wet sand, $\mathrm{u^*_b}$ is the expression in Equation \ref{eq:bagnold}, w is the water content in percent by mass, and RH is the relative humidity in percent by pressure. When RH or water content increases, the threshold will increase accordingly. The results of Belly (1964) are shown in Fig. \ref{fig:rhu} (threshold RH) and Fig. \ref{fig:watercontentu} (threshold water content).

Iversen et al. (1976), Iversen and White (1982), and Greeley and Iversen (1985) added interparticle forces and aerodynamic lift into the force balance and expanded the threshold model to small grains \textless200 $\mathrm{\mu}$m. Their model is a piecewise function in three Reynolds number regimes. On the basis of a more explicit expression of interparticle forces, Shao and Lu (2000) simplified the model of Greeley and Iversen (1985) to a single equation:
\begin{equation}
\label{eq:shaolu}
u^*_{sl}=\sqrt{f(Re^*)(\frac{\rho_p-\rho_a}{\rho_a}gd+\frac{\gamma}{\rho_ad})},
\end{equation}
with $\mathrm{f(Re^*)}$ approximately equal to 0.0123, and $\mathrm{\gamma}$ between 1.65$\mathrm{\times10^{-4}\ N\ m^{-1}}$ and 5$\mathrm{\times10^{-4}\ N\ m^{-1}}$. This threshold model is for loosely packed dry materials.

McKenna Neuman (2003) and McKenna Neuman and Sanderson (2008) slightly modified Shao and Lu's (2000) model for potentially humid or high/low temperature environments:
\begin{equation}
\label{eq:mckenna}
u^*_w=\sqrt{f(Re^*)(\frac{\rho_p-\rho_a}{\rho_a}gd+\frac{\gamma'}{\rho_ad^2})},
\end{equation}
where
\begin{equation}
\label{eq:gamma}
\gamma'=\frac{6}{\pi}\frac{a_i}{a_l}F_i,
\end{equation}
and
\begin{equation}
\label{eq:fi}
F_i=\beta_cd+|\Psi|A_c.
\end{equation}
The term $a_i/a_l$ is the ratio of the moment arm lengths of interparticle and lift forces (see Fig. \ref{fig:forcebalance}). The term $\beta_cd$ expresses the electrostatic and van der Waals forces. The term $|\Psi|A_c$ describes the effect of cohesion. $\Psi$ is the matric potential (also called Laplace pressure, $\mathrm{\Delta p}$) that describes the pressure difference caused by surface tension, and it can be expressed by the Kelvin equation,
\begin{equation}
\label{eq:psi}
\Psi=(\frac{RT}{V_l})\mathrm{ln}(\frac{\mathrm{RH}}{100}),
\end{equation}
where R=8.314 $\mathrm{J \ mol^{-1}\ K^{-1}}$ is the ideal gas constant for dry air, T is temperature in K, and $\mathrm{V_l}$ is the molar volume of the liquid (for water, $\mathrm{V_l}$=$\mathrm{1.8\times10^{-5} \ m^3\  mol^{-1}}$). $\mathrm{A_c}$ is the total contact area of adsorbed water films between particles and is approximated in McKenna Neuman and  Sanderson (2008) as,
\begin{equation}
\label{eq:ac}
A_c=\delta \pi kd(\delta/\delta_0)^n.
\end{equation}
$\delta_0$ is the thickness for a monolayer of adsorbed water, and $\delta$ is the water film thickness,
\begin{equation}
\label{eq:thickness}
\delta={(\frac{H}{6\pi\Psi})}^{1/3},
\end{equation}
and H is the Hamaker constant, an interaction parameter for adhesive surfaces ($\mathrm{-1.9\times10^{-19}}$ J, Iwamatsu and Horii, 1996; Tuller and Or, 2005). k is a dimensionless number describing the surface roughness ($\sim$ $\mathrm{10^{-4}}$--$\mathrm{10^{-5}}$), and the power n varies between 6--8, depending on the surface roughness and particle packing arrangement. Both k and n are determined by fitting the experimental data to the model. Thus threshold wind speed is a function of matric potential $\mathrm{\Psi}$. The introduction of matric potential is useful for both humid coastal areas and cold regions (McKenna Neuman and Nicklings, 1989, data shown in Fig. \ref{fig:watercontentu}) since it incorporates two variables, temperature and relative humidity,  into one single variable. However, its applicability to low density materials has never been tested. Based on Equation (4), the results for the threshold RH variation are shown in Fig. \ref{fig:rhu} for 125 $\mathrm{\mu}$m (k=$2.1\times\mathrm{10^{-4}}$, n=6.1) and 210 $\mathrm{\mu}$m quartz sand (k=$2.1\times\mathrm{10^{-4}}$, n=5.0). Note that compared to slope of the threshold RH variation for 400 $\mathrm{\mu}$m sand (Belly, 1964), the slope for 125 $\mathrm{\mu}$m and 210 $\mathrm{\mu}$m is lower, while it should be higher for smaller sediments according to Equation \ref{eq:mckenna}. This discrepancy may be due to the use of different threshold definitions (Fecan et al., 1999); Belly (1964) defined threshold as the point when bed movement is fully sustained, while McKenna Neuman and Sanderson (2008) defined it as the initiation of bed movement.

Fecan et al. (1999) and Ravi et al. (2004, 2006) further investigated the effect of humidity on soils with different amounts of clay. They argued that with the clay component in soil, the matric potential of McKenna Neuman and Nicklings (1989) for sand was no longer applicable because clay has much stronger adsorption forces to bond a layer of water film than quartz sand. Fecan et al. (1999) combined previous studies (Belly, 1964; Bisal and Hsieh, 1966; McKenna Neuman and Nickling, 1989; Saleh and Fryrear, 1995; Chen et al., 1996) and found an empirical formula for threshold as a function of gravimetric water content:
\begin{equation}
\label{eq:fecan}
\begin{aligned}
&\frac{u^*_w}{u^*_d} = 1 \ \mathrm{when} \ w<w'\\
&\frac{u^*_w}{u^*_d} = [1+1.21(w-w')^{0.68}]^{0.5} \ \mathrm{when} \ w>w'\\
&w'=0.0014(\%\mathrm{clay})^2+0.17(\%\mathrm{clay})
\end{aligned}
\end{equation}
where w is the water content per mass, and $u^*_w$ and $u^*_d$ respectively correspond to wet and dry threshold wind speeds. The result is shown in Fig. \ref{fig:watercontentu}, and can be compared with the data and fitting of Belly (1964).

Fecan et al. (1999) defined an initiation water content w', where they showed that once the water content of soil exceeds w', the threshold increases with increasing water content in soil. However, wind tunnel runs with clay and sandy loam under a range of humidities found different results. Ravi et al. (2006) found threshold increases with increasing RH only when RH is less than 40\% or greater than 65\%. When RH is between 40\% and 65\%, threshold decreased with increasing RH. The wind tunnel results data from Ravi et al. (2006) for two kinds of soil (different clay content) are shown in Fig. \ref{fig:rhu}. They explain the results as follows: 1) for low RH (RH\textless40\%), an adsorption layer covers the particle (which happens only for soil with a clay component) and the cohesion forces are dominated by the adsorption forces; 2) for high RH (RH\textgreater65\%), water condenses and forms liquid bridges between particles and the cohesion forces are mainly the capillary forces between liquid bridges; 3) for RH in between 40\% and 65\%, a transition between the adsorption forces and capillary forces occurs, resulting in lower interparticle forces (see Equation \ref{eq:acravi} below). They thus modified the interparticle forces $\mathrm{F_i}$ (Equation \ref{eq:fi}) in McKenna-Neuman (2003) by modifying the total contact area $\mathrm{A_c}$ to describe the transition region (45\%\textless RH\textless 65\%):
\begin{equation}
\label{eq:acravi}
A_c=\pi(\frac{w'}{\rho_w\sigma}-\frac{y}{2})d
\end{equation}
where w' is the soil moisture content, $\mathrm{\rho_w}$ is the water density, and y is the distance between the two contacting sphere particles. Because w' varies as $\mathrm{c|\Psi|^{-b}}$, and $\mathrm{b<1}$, the total cohesion $\mathrm{|\Psi|A_c}$ is proportional to $\mathrm{|\Psi|^{1-b}}$, and thus when RH increases, the cohesion forces decrease, leading to decreasing threshold wind speed.

Overall, previous studies of the effect of water on the threshold wind speeds agree that for coarse-grained materials, the more water the materials have, the higher the threshold. That is, when RH or water content of sand increases, the interparticle cohesion between the particles increases, leading to a higher threshold. However, for sand with clay components, the threshold decreases with an increase in RH when RH is between 45\% and 60\%, although it is still larger than threshold in dry conditions with dry materials (Ravi et al., 2006, see Fig. \ref{fig:rhu}). Thus, we should expect a higher threshold for low density materials when RH is high or when materials are not dried and thus have a high water content.

\section{Methods}
\subsection{Materials}
Special care must be taken for planetary wind tunnels to reproduce relevant environmental conditions. As described in the previous section and shown in Table \ref{table:planetarycondition}, there are several major planetary conditions that affect aeolian transportation: 1) transporting materials, 2) gravity, 3) atmospheric density, 4) atmospheric viscosity, and 5) density ratio. 

The materials transported by wind on inner Solar System terrestrial planets (Earth, Venus, and Mars) are mainly from silicate rock, with Earth sediments dominated by quartz sand and Venus and Mars sediments are mainly mafic basaltic sand. For materials that have been used in the TWT, both quartz sand (including white silica sand and beach sand from Cemexusa) and basaltic sand (acquired from Pisgah Crater) are easy to acquire and resemble the real aeolian sediments for Earth, Mars, and Venus. On the other hand, for icy worlds in the outer solar system, including Titan, Triton, and Pluto, the sediments are mainly organics. Analogs to those organic materials can be made in the laboratory (`tholins', Sagan et al., 1979 and Cable et al., 2012), but low yields and toxic composition means that they are not ideal for wind tunnel experiments, for which larger quantities are required ($\sim$ 3000 cm\textsuperscript{3} for TWT). Tholins may also behave differently room temperature than under Titan conditions. Laboratory experiments indicate that Titan tholins have an effective density ($\mathrm{\rho_{eff}}$) of 500--1100 kg/m\textsuperscript{3} (H\"orst and Tolbert, 2013) and material density ($\mathrm{\rho_m}$) of 1300--1400 kg/m\textsuperscript{3} (Imanaka et al., 2012). Effective density and material density are related by a shape and porosity factor (S). When the particles are perfect spheres without pores (S=1), the effective density and material density are equal; irregularities and porosity both decrease S (H\"orst and Tolbert, 2013). Here we use these two measurements to estimate the maximum and minimum values of the material density on icy bodies, including Titan.

To investigate aeolian planetary processes in an Earth laboratory, we have to use materials with lower densities to compensate for the higher gravity on Earth. For example, on Mars the material transported is basaltic sand with density of 3000 kg/m\textsuperscript{3}, but since Martian gravity is only about 3/8ths that of Earth, previous experiments have used lower density material (1100 kg/m\textsuperscript{3}) to simulate the weight of the materials as transported on Mars. Table \ref{table:planetarycondition} shows the density for equivalent weight aeolian materials on other planetary bodies. Low density materials that have been used in previous wind tunnel experiments (Greeley et al., 1980; Burr et al., 2015a) and which are investigated here are walnut shells (from Eco-shell, Inc), gas chromatograph packing materials made from flux calcined diatomite (GC tan, Johns-Manville), iced tea powder (4C Totally Light), instant coffee (Foodhold U.S.A., LLC), activated charcoal (Sigma-Aldrich), and glass bubbles (3M).

To extend the previous work into threshold conditions on Titan (Burr et al., 2015a), we include additional materials with different densities. These additional materials include non-acid washed and acid washed glass beads (Mo-Sci Corporation, Sigma-Aldrich, 2500 kg/m\textsuperscript{3}),  gas chromatograph packing materials made from calcined diatomite (GC pink, Johns-Manville, 2150 kg/m\textsuperscript{3}), and chromite (Reade Advanced Materials, 4000 kg/m\textsuperscript{3}). The materials investigated in this work include all the previously and currently used TWT materials (Greeley et al., 1980; Burr et al., 2015a). The materials are summarized in Table \ref{table:material} in order of decreasing literature density values. The materials investigated in this study are the same batches (except iced tea powder and instant coffee) as the ones at the TWT, thus having the same size range as well as composition.

\setlength\LTleft{-0.8in}
\setlength\LTright{-0.8in}
\begin{longtable}{|c|c|c|c|c|c|c|}
\captionsetup{width=10cm}
\caption{Summary of planetary conditions. Values for Venus, Earth, Mars, and Titan are adopted from Burr et al. (2015b). For Triton and Pluto, atmospheric density values are derived using the ideal gas law, and surface temperature and pressure are adopted from Smith et al. (1989) and Gladstone et al. (2016), respectively. The atmospheric viscosity for Triton and Pluto is calculated by using gas type and temperature at http://www.lmnoeng.com/Flow/GasViscosity.htm.} \label{table:planetarycondition} \tabularnewline
\endfirsthead
\hline
Planetary & Density of & Gravity & Density of  & Atmospheric & Atmospheric &  Density \\  
Body & Material & (m/s\textsuperscript{2}) & Equivalent & Density & Viscosity & Ratio \\
& $\rho_p$ (kg/m\textsuperscript{3}) & & Weight Material & $\rho_a$ (kg/m\textsuperscript{3}) & (Pa$\cdot$s) & ($\rho_p$/$\rho_a$)\\
& & & on Earth (kg/m\textsuperscript{3}) & & &\\
  \hline\hline
  \multirow{2}{*}{Venus} & 3000 & \multirow{2}{*}{8.9} & \multirow{2}{*}{2724} & \multirow{2}{*}{65} & \multirow{2}{*}{$\mathrm{3.27\times10^{-2}}$} & \multirow{2}{*}{$\mathrm{46}$}\\
 & basalt & & & & &  \\
\hline
\multirow{2}{*}{Earth} & 2650 & \multirow{2}{*}{9.8} & \multirow{2}{*}{2650} & \multirow{2}{*}{1.2} & \multirow{2}{*}{$\mathrm{1.85\times10^{-5}}$} & \multirow{2}{*}{$\mathrm{2.2\times10^3}$}\\
 & quartz & & & & &  \\
 \hline
 \multirow{2}{*}{Mars} & 3000 & \multirow{2}{*}{3.7}  & \multirow{2}{*}{1132} & \multirow{2}{*}{0.015} & \multirow{2}{*}{$\mathrm{1.30\times10^{-5}}$} & \multirow{2}{*}{$\mathrm{2\times10^5}$}\\ 
 & basalt & & & & & \\
\hline
\multirow{2}{*}{Titan} & 500--1400& \multirow{2}{*}{1.4} & \multirow{2}{*}{71--200} & \multirow{2}{*}{5.1} & \multirow{2}{*}{$\mathrm{6.25\times10^{-6}}$} & \multirow{2}{*}{78--294}\\
 & organics & & & & & \\
\hline
\multirow{2}{*}{Triton} & 500--1400 & \multirow{2}{*}{0.62} & \multirow{2}{*}{31--89} & \multirow{2}{*}{$\mathrm{\sim 9\times10^{-5}}$} & \multirow{2}{*}{$\mathrm{\sim 2\times10^{-6}}$} & \multirow{2}{*}{3.1--12.0$\mathrm{\times10^6}$}\\
& organics& & & & & \\
\hline
\multirow{2}{*}{Pluto} & 500--1400 & \multirow{2}{*}{0.78} & \multirow{2}{*}{40--111} & \multirow{2}{*}{$\mathrm{\sim 9\times10^{-5}}$} & \multirow{2}{*}{$\mathrm{\sim 2\times10^{-6}}$} & \multirow{2}{*}{3.1--12.0$\mathrm{\times10^6}$}\\
& organics& & & & & \\\hline
\end{longtable}

\begin{table}[h!]
\centering
\caption{Summary of material properties. GC indicates Gas Chromatograph packing materials. GC tan is calcined diatomite: according to Burr et al. (2015a), it a has different color compared to GC pink. For the literature density values, chromite, basalt, quartz sand, beach sand, and glass beads are standard values. Density of the GC pink, GC tan, activated charcoal, and glass bubbles were provided by the manufacturer. Density of walnut shells is originated from Greeley et al. 1980. Density of iced tea and instant coffee comes from FAO/INFOODS Density Database.}
\vspace{0.2cm}
 \begin{tabular}{|c|c|c|} 
 \hline
Material Name & Density in  & Size Range  \\  
 & Literature (kg/m\textsuperscript{3}) & ($\mu$m) \\
  \hline\hline
  Chromite & 4000 & 212--250; 250--300\\
   \hline
   Basalt & 3000 & 150--250; 250--500; 707--1000 \\
      \hline
  \multirow{2}{*}{Quartz Sand} &\multirow{2}{*}{2650} & 106--125; 125--150; 150--175; \\
 & & 175--212; 212--250 \\
    \hline
  \multirow{2}{*}{Beach Sand} & \multirow{2}{*}{2650} &  500--600; 600--700;\\
  & & 707--833; 833--1000 \\
     \hline
 \multirow{3}{*}{Glass Beads} & \multirow{3}{*}{2500} &  150--180 (non-acid washed);\\
 & & 150--180 (acid washed);\\
  & & 180--212; 500--600 \\
     \hline
  GC pink & 2150 & 125--150\\
\hline
  GC tan & 1300 & 150--175 \\
     \hline
 \multirow{2}{*}{Walnut Shells} & \multirow{2}{*}{1100} & 125--150; 150--175; 175--250;\\ 
  & & 500--600; 707--833; 833--1000\\ 
     \hline
  Iced Tea Powder& 1030 & N/A \\ 
     \hline
  Activated Charcoal & 400 & 250--300; 425--500; 600--707\\
     \hline
 Instant Coffee & 250 & N/A \\ 
    \hline
  Glass Bubbles & 100--140 & 30--115 \\
   \hline
\end{tabular}
\vspace{-0.2cm}
\label{table:material}
\end{table}

\subsection{Density Measurements}

Measuring particle density requires a series of careful measurements. Mass is straightforward to obtain with an analytical balance. For this work, the mass of the materials was measured by an analytical balance (Satorius Entris 224-1S), with standard deviation of 0.1 mg. 

\begin{figure}
\begin{center}
\includegraphics[width=\textwidth]{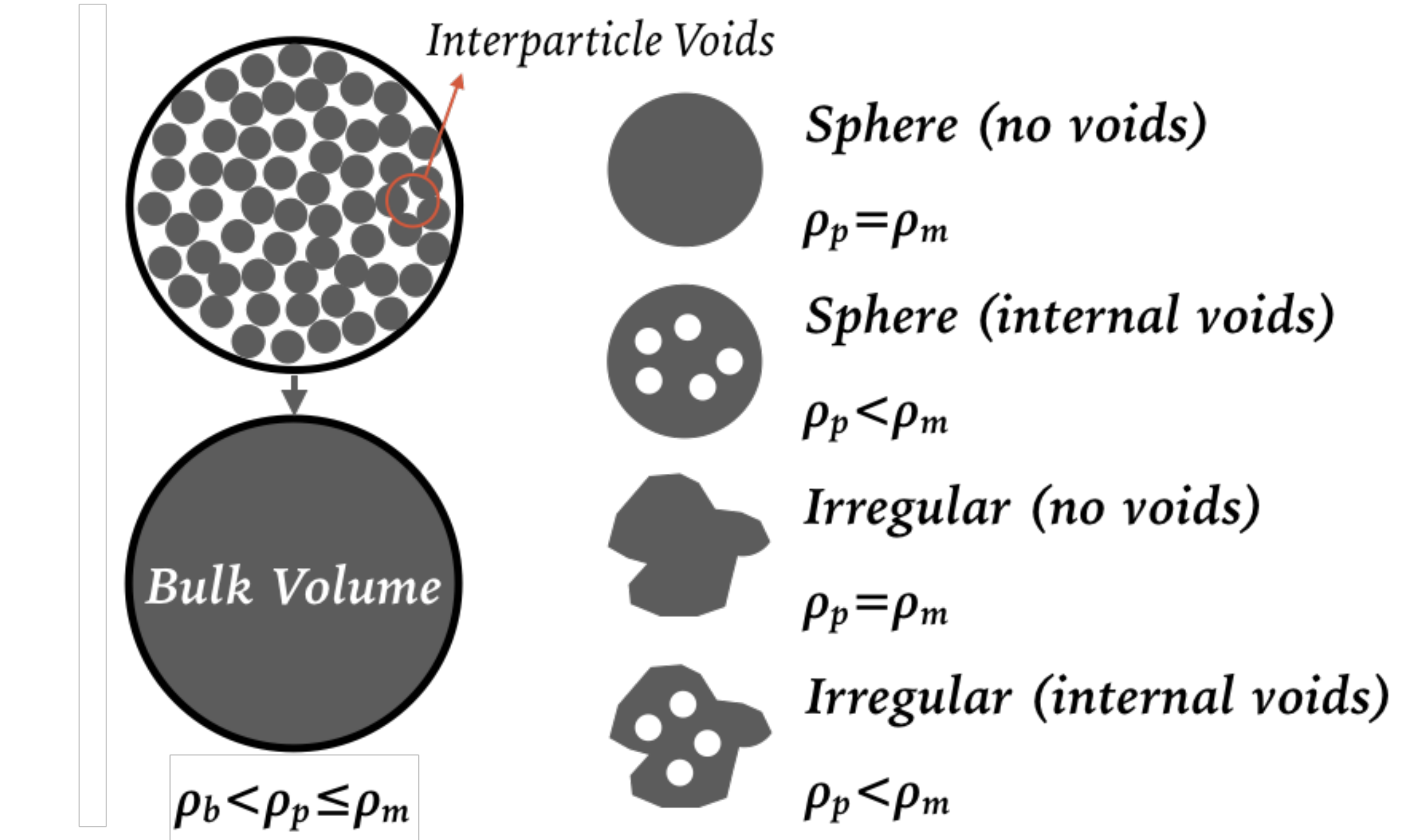}
\end{center}
\caption[]{Comparison of different densities: bulk density ($\mathrm{\rho_b}$), particle density ($\mathrm{\rho_p}$), and material density ($\mathrm{\rho_m}$), adapted from Webb (2001) and H{\"o}rst and Tolbert (2013).}
\label{fig:density}
\end{figure}

The particle volume is more difficult to measure, because of a number of different definitions of density. Fig. \ref{fig:density} compares three densities: the bulk density ($\mathrm{\rho_b}$), particle density ($\mathrm{\rho_p}$), and material density ($\mathrm{\rho_m}$). Bulk density has the smallest value, as the bulk volume includes: 1) volume of the solid material, 2) closed internal voids, 3) open pores of particles, and 4) interparticle voids. The volume defined in particle density ($\mathrm{\rho_p}$) includes the volume of the solid material and the volume of the internal closed pores, whereas the material volume ($\mathrm{\rho_m}$) only includes the solid material volume. The particle density can be smaller or equal to the material density, depending on porosity. When the particles have internal pores, the particle density is always smaller than the material density (Fig. \ref{fig:density}). Conversely, when particles have no pores, the particle density is equal to the material density (Webb, 2001, also see Fig. \ref{fig:density}).

\begin{figure}
\begin{center}
\includegraphics[width=\textwidth]{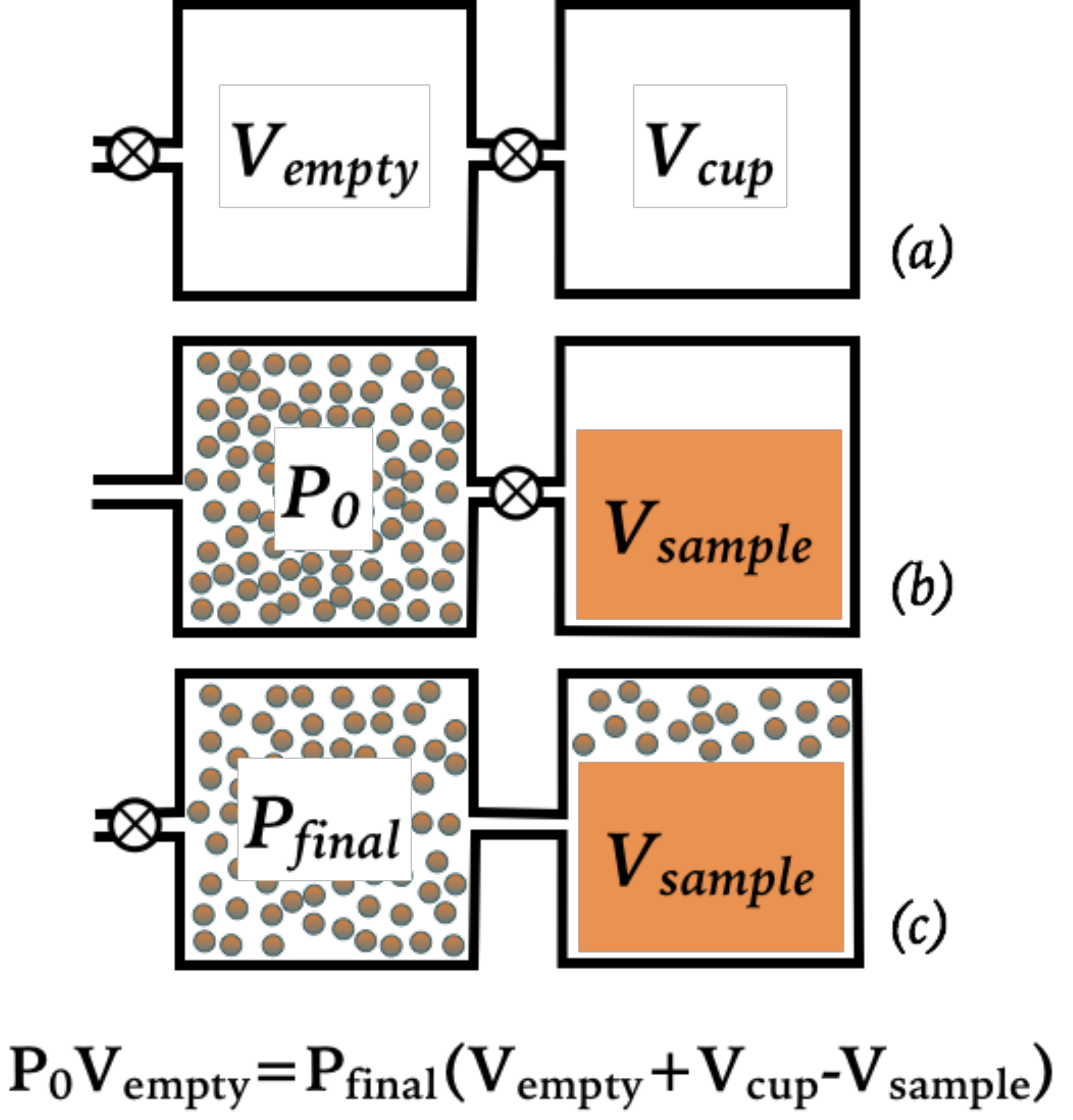}
\caption[]{Pycnometer work flow and the ideal gas law used to calculate the volume of the sample ($\mathrm{V_{sample}}$).}
\label{fig:pycnometer}
\end{center}
\end{figure}

Here we used an AccPyc II 1340 Automatic Gas (Helium) Pycnometer to measure the volume of the materials. The principle of the pycnometer is the gas displacement method and is illustrated in Fig. \ref{fig:pycnometer}. Helium gas is first admitted into an empty compartment with calibrated volume $\mathrm{V_{empty}}$, until it equilibrates with a certain pressure (Fig. \ref{fig:pycnometer}(b)). The samples are sealed in a second calibrated cup with volume $\mathrm{V_{cup}}$. After the pressure is stable in the empty compartment ($\mathrm{P_0}$), the helium gas is discharged from the empty compartment to the cup with the samples. The helium gas fills the spaces within the sample as small as $\sim$3 $\mathrm{\AA}$ rapidly, and the final equlibrated pressure in the system is recorded as $\mathrm{P_{final}}$ (Fig. \ref{fig:pycnometer}). Using the ideal gas law:
\begin{equation}
\label{eq:density}
P_0V_{\mathrm{empty}}=P_{\mathrm{final}}(V_{\mathrm{empty}}+V_{\mathrm{cup}}-V_{\mathrm{sample}}),
\end{equation}
the pycnometer calculates the volume of the sample, $\mathrm{V_{sample}}$. The volume of the materials (V) and the standard deviation are given automatically by the pycnometer after 10 purges and 20 runs with the materials.

\subsection{Gravimetric Water Content Measurements}
We determine the water content of the materials by gravimetric measurements. The Relative Humidity (RH) and temperature in laboratory were recorded using a digital hygrometer (Dwyer Instrument) with 0--100\% RH range (accuracy of $\pm$2\%) and $-$30--85$\degree$C (accuracy $\pm$ 0.5$\degree$C). The materials were put in an aluminum foil boat during the measurement. To eliminate the water adsorption of the aluminum foil, we put the foil in a 120$\degree$C oven (Lab Safety Supply Model No.32EZ28, temperature accuracy $\mathrm{\pm1\degree C}$ at 100$\degree$C) for 24 hours, and weighed it immediately after removal ($\mathrm{m^{'}_{dry}}$). The weight of the aluminum foil boat increases over time until equilibrating with water moisture in the atmosphere ($\mathrm{m^{'}_{wet}}$), usually in about 10 minutes. Thus, the amount of water adsorption on the aluminum foil is:
\begin{equation}
\label{eq:watercontent}
\Delta m^{'}=m^{'}_{wet}-m^{'}_{dry}
\end{equation}
After the equilibration of the aluminum foil boat, we laid a thin layer of materials on the bottom of the boat. The materials and the boat were then dried together in the 120$\degree$C oven for 24 hours. A lower temperature (105 $\degree$C) was tried to bake the materials, but it didn't change the overall results. After drying, they were weighed again immediately ($\mathrm{m_{dry}}$). Then we left the materials in air to let them equilibrate, weighing them every 0.5--5 minutes. When the weight of the materials no longer changed with time, we recorded this final weight ($\mathrm{m_{wet}}$), ambient RH, temperature, and the time the materials took to equilibrate ($t_{eq}$). The final water content of the materials after they equilibrate (at a given RH and temperature) is given by:
\begin{equation}
\label{eq:watercontent2}
u(\%)=\frac{m_{wet}-m_{dry}-\Delta m^{'}}{m_{dry}-m^{'}_{dry}}.
\end{equation}

\subsection{Thermogravimetric (TGA) Measurements}

Since many of the low density materials we used are porous, water is both adsorbed on the surface and absorbed in the interior of the particles. The surface water affects threshold by increasing interparticle cohesion, whereas the interior water changes the density of the materials. Therefore it is important to differentiate between surface and internal water. Surface and internal water are released at different temperatures and can be separated using thermogravimetric analysis (TGA). The samples, weighing 10--50 mg, were placed in an aluminum crucible and then loaded into a Mettler TGA/SDTA851e purged with nitrogen. The samples were heated from 25.0 to 600.0$\degree$C at a rate of 10.00$\degree$C/min. A slower heating rate (5.00$\degree$C/min) was tested on a walnut shell sample, but it did not change the overall results. The RH during the experiment was measured by the RH probe described in Section 2.3.

\subsection{Titan Wind Tunnel Experiments Using `Wet' and `Dry' Sediments}
To experimentally investigate the effect of water adsorption on threshold, we ran a set of experiments in the TWT at a range of pressures as a comparison study of walnut shells (size 125--150 $\mu$m) that were either subject to drying (`dry') or were in equilibrium with the ambient humidity at 1 bar (`wet'). The small end of the sediment size (125--150 $\mathrm{\mu m}$) was chosen for the `wet' and `dry' runs because smaller particles are more sensitive to interparticle force change than larger particles with greater gravitational forces. To measure their water content, the `wet' walnut shells were analyzed after the TWT run following the method described in Section 3.3, using a different analytical balance (A\&D HR-120 with standard deviation of 0.1 mg) and oven (VWR Economy Vacuum Oven Model 1400E, temperature accuracy $\mathrm{\pm3.5\degree C}$ at 100$\degree$C). 

To prepare the `dry' walnut shells, we put the materials needed for a TWT run (approximately 3000 cm\textsuperscript{3}) in a 120$\degree$C oven for 24 hrs. Then we transferred all the materials into a desiccator (Lab Safety Supply, I.D. 300mm) with desiccant (Carolina, Silica Gel, Indicating Beads, Laboratory Grade) in preparation for the TWT experiment. While we set the bed for the TWT experiment (see Extended Data Figure 2 in Burr et al. 2015a), the materials were exposed to ambient air for 40 minutes. We chose the walnut shells for this experiment because their equilibration timescale, as discussed in Section 4.2, is longer than the time required to set the bed. The procedure for conducting experiments in the TWT can be found in Burr et al. (2015a). For the experiments presented here, the pressures in the TWT were 1, 3, 8, 12.5, 15, and 20 bars. The freestream wind speed was converted from dynamic pressure collected by pitot tubes in the TWT (for details, see Methods in Burr et al., 2015a; the only change is that the current TWT has the fixed pitot tube in the test section to collect dynamic pressure, instead of at the back of wind tunnel as described in Burr et al., 2015a).

\section{Results and Discussion}

\subsection{Particle density measurement of wind tunnel materials}

The particle density measurements show that for materials with densities over 2000 $\mathrm{kg/m^3}$ (chromite, basalt, quartz sand, beach sand, glass beads, and GC pink), the measured densities are very close to the densities reported in the literature (see Table \ref{table:density1}). However, for material densities less than 2000 $\mathrm{kg/m^3}$ (GC tan, walnut shells, instant coffee, activated charcoal, iced tea powder, and glass bubbles), the measured densities differ from those in the literature or as provided by the manufacturer (see Table \ref{table:density2}). Thus here we can divide the materials into two groups, high density and low density materials, where the division between low density and high density materials is 2000 $\mathrm{kg/m^3}$.

The discrepancy between the particle density measured by the helium gas pycnomter and the density reported in the literature or by the manufacturer could be attributed to the different density definitions. The helium gas in the pycnometer can rapidly fill the open pores of the materials, thus the pycnometer measures the particle density ($\mathrm{\rho_p \textless \rho_m}$ if the particles have closed internal pores, or $\mathrm{\rho_p=\rho_m}$ if the particles have no internal pores). The density reported in the literature may be bulk density given by the manufacturer, as it is with activated charcoal, iced tea powder, instant coffee, and GC tan (calcined diatomite). The density used for walnut shells in the literature is 1100 $\mathrm{kg/m^3}$ (Greeley et al., 1980; Burr et al., 2015a), whereas the density measured by pycnometer gives 1400 $\mathrm{kg/m^3}$. One possible explanation is that the densities given by the manufacturer are defined in other ways or are not measured precisely. The density measured for the high density materials are likely closer to the literature value because those materials generally have no internal voids. However, the `density' used in the TWT data analysis depends also on the porosity, surface area, size, and shape of the particles. Therefore, this value should fall between the bulk density and the material density, but probably closer to the material density because the wind can penetrate the interparticle voids.
\setlength\LTleft{-0.1in}
\setlength\LTright{-0.1in}
\begin{longtable}{|c|c|c|c|c|}
\captionsetup{width=10cm}
\caption{Summary of densities of high density wind tunnel materials (literature density greater than 2000 $\mathrm{kg/m^3}$) in literature and measured by the pycnometer, with standard deviation in the measurements.}\label{table:density1} \tabularnewline
\endfirsthead
 \hline
 Material & Size Range  & Density in   & Updated  &  Standard   \\  
Name & ($\mu$m)  & Literature & Density & Deviation \\
 & & (kg/m\textsuperscript{3}) & (kg/m\textsuperscript{3}) & (kg/m\textsuperscript{3}) \\
\hline\hline
\multirow{2}{*}{Chromite} & 212--250 & \multirow{2}{*}{4000} & 4524.7 & 1.1\\
& 250--300 &  & 4518.8 & 1.1\\
\hline
\multirow{3}{*}{Basalt} & 150--250 & \multirow{3}{*}{3000} & 2841.4 & 1.1\\
 & 250--500 & & 2882.8 & 2.0\\
 & 707--1000 &  & 2919.0 & 1.0\\
\hline
Quartz Sand & 175--212 & 2650 & 2656.4 & 1.2 \\
\hline
\multirow{4}{*}{Beach Sand} & 500--600 & \multirow{4}{*}{2650} & 2631.2 & 2.2\\
 & 600--707 &  & 2639.0 & 0.8\\
 & 707--833 &   & 2634.3 & 1.2\\
 & 833--1000 &   & 2632.3 & 1.1\\
\hline
\multirow{4}{*}{Glass Beads} & 150--180 (non-acid washed) & \multirow{4}{*}{2500} & 2481.1 & 0.4\\
 & 150--180 (acid-washed) &  & 2420.2 & 0.6\\
 & 180--212 &   & 2634.3 & 1.2\\
 & 500--600 &   & 2632.3 & 1.1\\
\hline
GC pink & 125--150 & 2150 & 2364.6 & 2.2\\
\hline
\end{longtable}

\begin{table}[h!]
\centering
\caption{Summary of the densities of low density wind tunnel materials (literature density less than 2000 $\mathrm{kg/m^3}$) in the literature and measured by the pycnometer, with standard deviation in the measurements.}
\vspace{0.2cm}
\begin{tabular}{|c|c|c|c|c|} 
 \hline
 Material & Size  & Density in   & Updated  &  Standard   \\  
& Range & $\mathrm{Literature}$ & Density & Deviation \\
 &$\mu$m & (kg/m\textsuperscript{3}) & (kg/m\textsuperscript{3}) & (kg/m\textsuperscript{3}) \\
  \hline\hline
 GC tan & 150--175 & 1300 & 2006.1 & 5.0 \\
\hline
\multirow{6}{*}{Walnut Shells} & 125--150 & \multirow{6}{*}{1100} & 1426.5 & 1.0\\
& 150--175 &  & 1419.1 & 0.9\\
& 175--250 & & 1418.8 & 1.0\\
& 500--600 &  & 1415.4 & 2.6\\
& 707--833 &  & 1407.2 & 2.7\\
& 833--1000 &  & 1411.8 & 1.7\\
\hline
Iced Tea Powder & N/A & 1030 & 1423.5 & 1.3\\ 
\hline
\multirow{3}{*}{Activated Charcoal} & 250--300 & \multirow{3}{*}{400} & 1932.4 & 6.9\\
& 425--500 &  & 1896.7 & 15.9\\
& 600--707 &  & 1881.8 & 12.5\\
\hline
Instant Coffee & N/A & 250 & 1473.9& 1.2\\ 
\hline
Glass Bubbles & 30--115 & 100--140 & 140.1 & 0.6\\ 
\hline
\end{tabular}
\vspace{-0.2cm}
\label{table:density2}
\end{table}

\subsection{Water content and equilibration timescales of wind tunnel materials}
\begin{figure}
\begin{center}
\includegraphics[width=\textwidth]{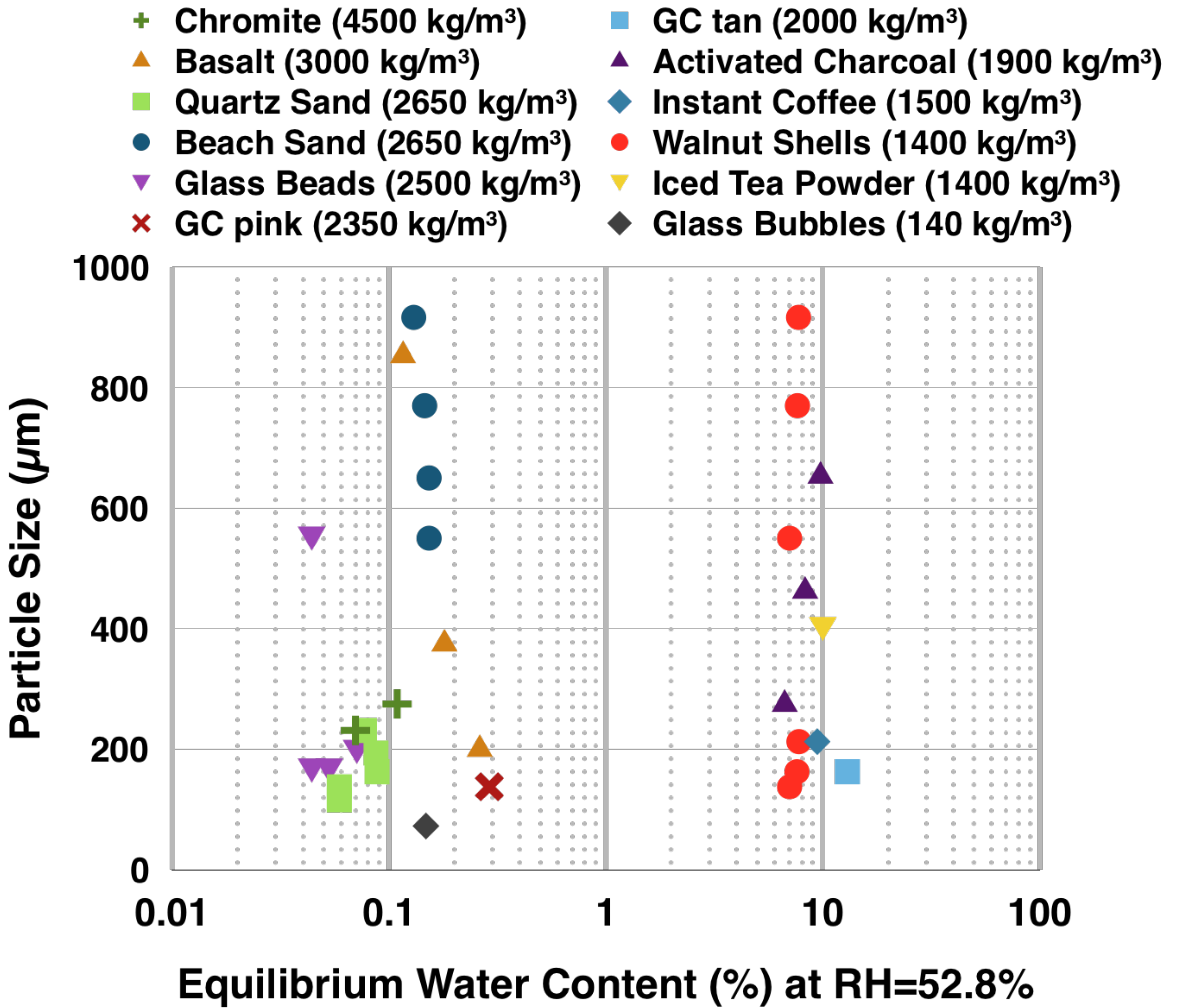}
\end{center}
\caption[]{Water content of wind tunnel materials of different sizes at the same RH (RH=52.8\%). Density values for the materials are adopted from the pycnometer measurements in Section 4.1.}
\label{fig:watercontent}
\end{figure}

\begin{figure}
\begin{center}
\includegraphics[width=\textwidth]{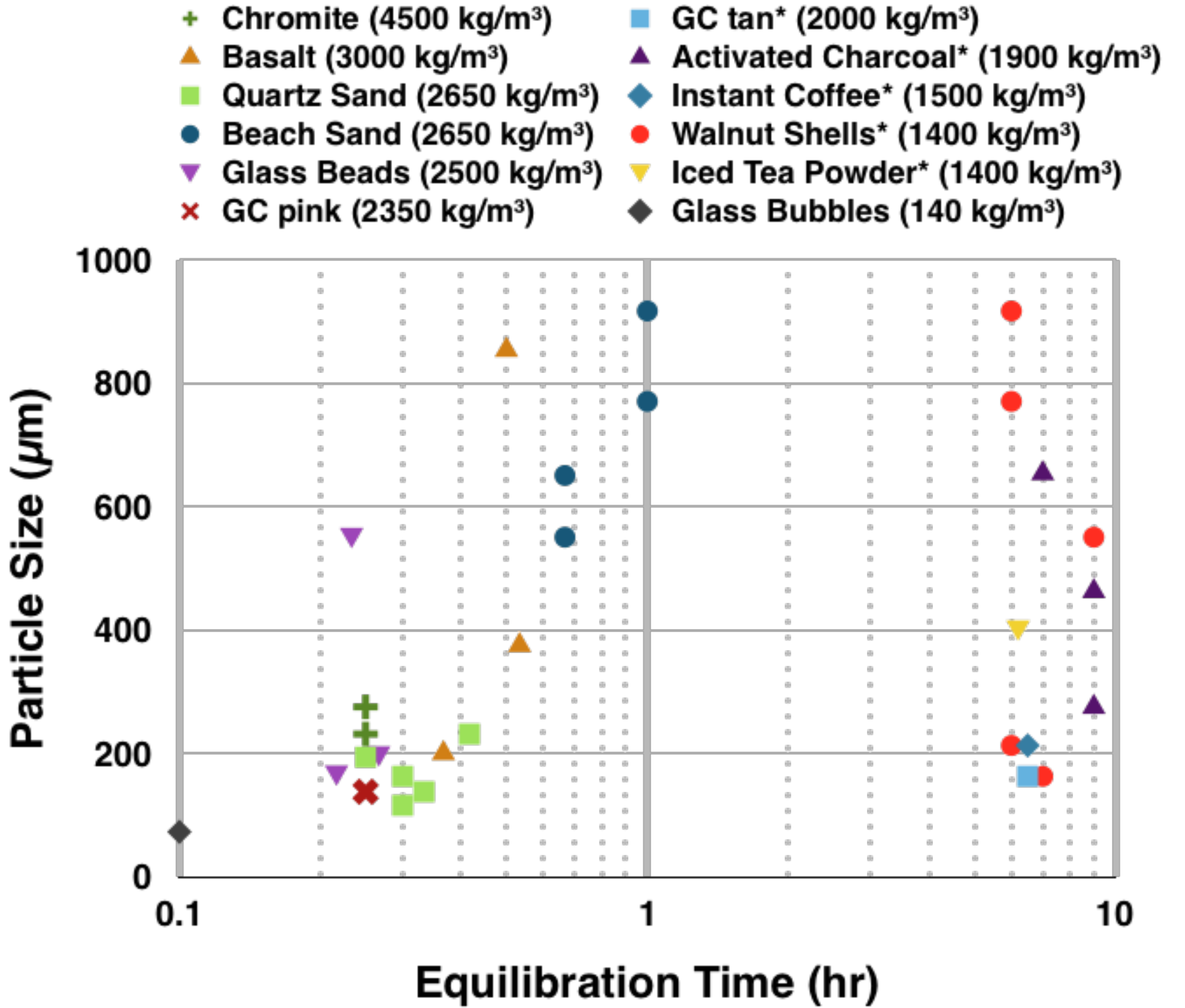}
\end{center}
\caption[]{Equilibration Timescales of wind tunnel materials of different sizes. For the materials marked with * (GC tan, activated charcoal, instant coffee, walnut shells, and iced tea powder), the equilibrium timescales were long, so that the minimum equilibration timescales are plotted. Density values for the materials are adopted from the pycnometer measurements in Section 4.1.}
\label{fig:equilibrationtime}
\end{figure}

The gravimetric measurements allow us to classify the materials by water content. As shown in Fig. \ref{fig:watercontent}, the materials can be divided into 2 groups: 1) materials with low water content (\textless 1\%), including all materials with literature densities over 2000 kg/m\textsuperscript{3} (high density materials) and glass bubbles and 2) materials with high water content (\textgreater 6\%), including materials with literature densities less than 2000 kg/m\textsuperscript{3} (low density materials), except glass bubbles. For the equilibration timescales shown in Fig. \ref{fig:equilibrationtime}, we can classify the materials in the same way: low water content materials have a short equilibration time, usually less than 1 hr, while high water content materials have a long equilibration time, over 6 hrs. For the same kind of material, both the water content and equilibration timescales show no apparent size dependence (Fig. \ref{fig:watercontent} and Fig. \ref{fig:equilibrationtime}).

One possible explanation for the high water content of low density materials is the combination of being hydrophilic and having large surface area-to-volume ratios, whereas the high density materials generally have smaller surface area-to-volume ratios and are hydrophobic. Glass bubbles are the exception; they are low density but they also have low water content ($\sim$0.05\%) and short equilibration time ($\sim$15 min) like high density materials. This is because they are designed to have low surface-area-to-volume ratio and are hydrophobic. 

\begin{figure}
\begin{center}
\includegraphics[width=\textwidth]{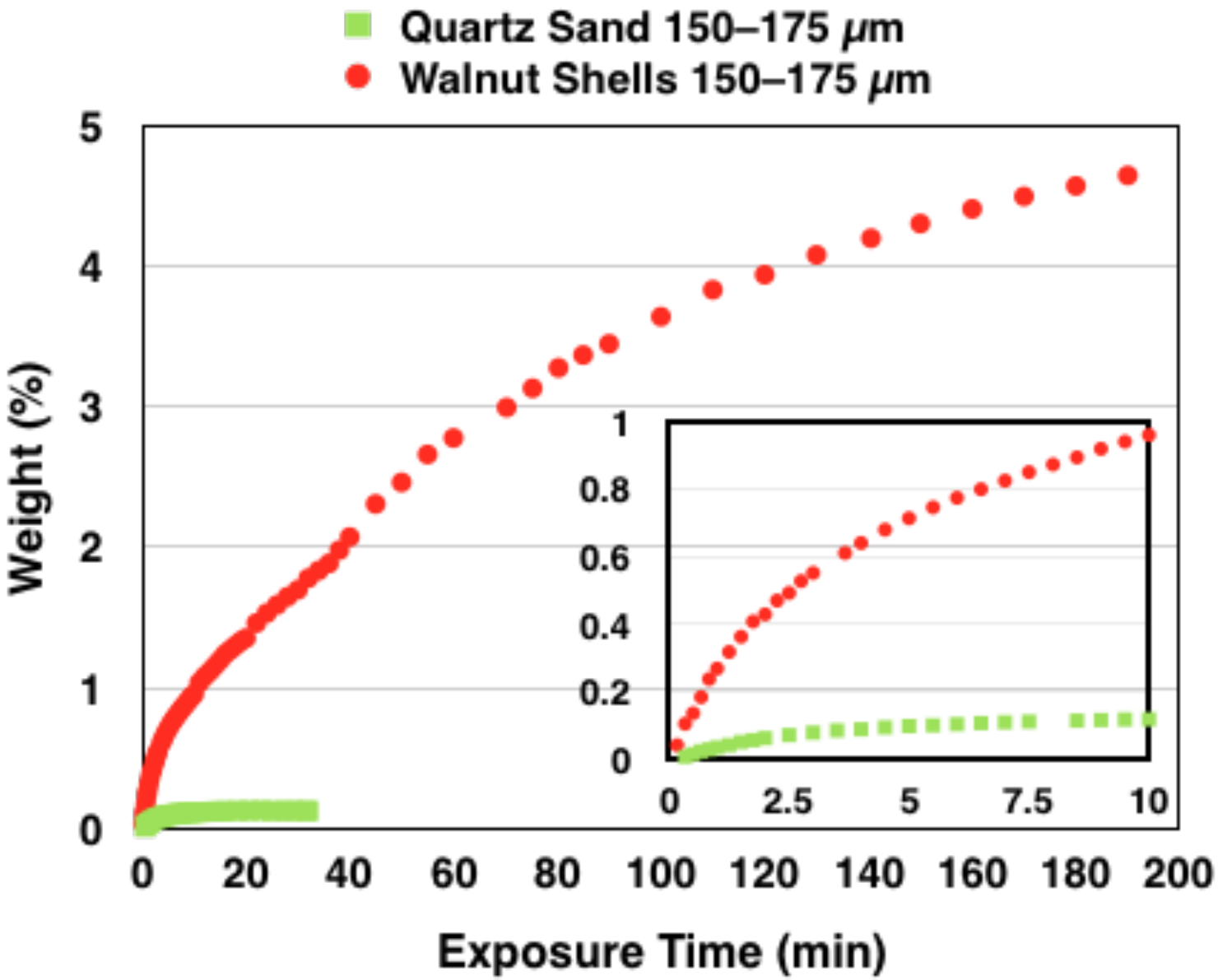}
\end{center}
\caption[]{Comparison of water content and equilibration process of one low density (walnut shells 150--175 $\mu$m) and one high density (quartz sand 150--175 $\mu$m) wind tunnel material up to 200 minutes. The inset graph magnifies the comparison in the first 10 minutes.}
\label{fig:comparison}
\end{figure}

In Fig. \ref{fig:comparison} we present the equilibration curves of two typical wind tunnel materials, quartz sand (low water content, high density) and walnut shells (high water content, low density). The walnut shells have much higher water content and equilibrate more slowly than quartz sand. In the first 10 minutes when quartz sand is approaching equilibrium, the walnut shells adsorb much more water by weight compared to quartz sand in the same time period. This correlation indicates water content and equilibration timescales are related. The long equilibration time for the low density materials also indicates that these materials cannot be dried quickly by circulating dry air in the wind tunnel. However, a short exposure time to air for the low density materials will not increase the water content to the equilibrium state.

We used the natural variation of humidity in the laboratory (15--60\%) to see how water content varies as a function of RH. There is a linear relationship between RH and water content for some of the materials shown in Table \ref{table:rhwatercontent}, including basalt, beach sand, walnut shells, activated charcoal, GC tan, iced tea powder, and instant coffee. They all have $\mathrm{R^2}$ values greater than 0.8 for a linear relationship. These linear relationships could be used to estimate water content of materials when only RH is recorded. For chromite, glass beads, quartz sand, GC pink, and glass bubbles, the coefficients of determination $\mathrm{R^2}$ for the linear relationships are only between 0.3--0.6. There may be a linear relationship between RH and water content for these materials as well, but the relationship is difficult to measure without a more precise analytical balance because of the low water content of these materials.

\begin{table}{}
\centering
\caption{Summary of the RH and water content linear relationship of the wind tunnel materials, with $\mathrm{R^2\textgreater0.8}$. The linear relationship is $\mathrm{y=ax+b}$, where y is the water content by mass and x is the RH in \%. $\mathrm{R^2}$ is the coefficient of determination for each linear relationship. Quartz sand (all sizes), GC pink (125--150 $\mu$m), and glass bubbles have $\mathrm{R^2}$ values that vary between 0.3--0.6, because their water content is very small ($<1\%$) compared to other materials and the measured water content values have large deviations. }
\vspace{0.2cm}
 \begin{tabular}{|c|c|c|c|c|} 
 \hline
 Material & Size Range ($\mu$m) & a & b & $\mathrm{R^2}$  \\  
  \hline\hline
 \multirow{3}{*}{Basalt} & 150--250 & $\mathrm{1.93\times10^{-3}}$ & $\mathrm{1.60\times10^{-1}}$ & 0.944 \\
  & 250--500 & $\mathrm{1.38\times10^{-3}}$ & $\mathrm{1.06\times10^{-1}}$ & 0.954 \\
  & 707--1000 & $\mathrm{8.70\times10^{-4}}$ & $\mathrm{6.58\times10^{-2}}$ & 0.941\\
 \hline
 \multirow{4}{*}{Beach Sand} & 500--600 & $\mathrm{1.22\times10^{-3}}$ & $\mathrm{8.62\times10^{-2}}$ & 0.930\\
& 600--700 & $\mathrm{1.17\times10^{-3}}$ & $\mathrm{8.87\times10^{-2}}$ & 0.935\\
& 707--833 & $\mathrm{1.15\times10^{-3}}$ & $\mathrm{8.34\times10^{-2}}$ & 0.964\\
& 833--1000 & $\mathrm{1.06\times10^{-3}}$ & $\mathrm{7.13\times10^{-2}}$ & 0.953\\
 \hline
 \multirow{6}{*}{Walnut Shells} & 125--150 & $\mathrm{9.87\times10^{-2}}$ & 2.07 & 0.898\\ 
 & 150--175 & $\mathrm{8.28\times10^{-2}}$ & 3.26 & 0.850\\ 
 & 175--250 & $\mathrm{8.24\times10^{-2}}$ & 3.41 & 0.835\\ 
 & 500--600 & $\mathrm{8.02\times10^{-2}}$ & 3.28 & 0.794\\
 & 707--833 & $\mathrm{7.90\times10^{-2}}$ & 3.58 & 0.903 \\
 & 833--1000 & $\mathrm{8.04\times10^{-2}}$ & 3.51 & 0.815 \\
 \hline
 Activated Charcoal & 400--841 & $\mathrm{2.30\times10^{-1}}$ &$-\mathrm{3.05\times10^{-1}}$ & 0.946\\
 \hline
 GC tan & 150--175 & $\mathrm{1.02\times10^{-3}}$ & $\mathrm{7.88\times10^{-2}}$ & 0.927 \\
 \hline
 Iced Tea Powder& N/A & $\mathrm{1.13\times10^{-1}}$ & 3.79 & 0.929 \\ 
 \hline
 Instant Coffee & N/A & $\mathrm{1.13\times10^{-1}}$ & 4.32 & 0.875 \\ 
   \hline
\end{tabular}
\vspace{-0.2cm}
\label{table:rhwatercontent}
\end{table}

\subsection{Surface and internal water of wind tunnel materials}
As discussed in Section 3.4, surface water can change the interparticle cohesion and affect the threshold. The surface water measurements from TGA are listed in Table \ref{table:tga}.

For materials with less than 0.2\% water content, the mass loss was below the limit of detection for the TGA. Thus for high density materials like basalt, quartz sand, beach sand, and chromite, the surface and internal water content cannot be detected using the TGA. However, the high density materials are not porous or hydrophilic, thus the surface water content should equal the total water content, which we measured by gravimetric analysis.

For the low density materials like walnut shells, iced tea powder, and instant coffee, we can only get partial information from TGA, because thermal reactions will occur for these materials at high temperature. Generally, the surface water of a material releases from about 50$\degree$C to 150$\degree$C, then its internal water starts to release from about 200$\degree$C. Walnut shells start to release their internal water from about 175--200$\degree$C, but thermal destruction begins around 202$\degree$C (Findor{\'a}k et al., 2016), so we cannot get the internal water content of walnut shells directly from TGA measurement. For iced tea powder, thermal destruction happens at the lowest temperature of all the materials investigated, which is about 150$\degree$C. For instant coffee, thermal destruction happens at 175$\degree$C. Activated charcoal is stable during the entire heating process until 600$\degree$C. From Table \ref{table:tga}, we can find that surface water occupies over 80\% of the total water content for activated charcoal. Even for activated charcoal with extremely high porosity, the surface water still dominates. 

The estimated total water content using the linear relationship in Table \ref{table:rhwatercontent} should equal the sum of surface water and internal water. However, it seems clear from this analysis that most of the water measured, if not all, by gravimetric analysis is surface water.

\setlength\LTleft{-0.4in}
\setlength\LTright{-0.4in}
 \begin{longtable}{|c|c|c|c|c|c|} 
\caption{Separation of surface and internal water from TGA analysis for some of wind tunnel materials. We calculated the estimated water content values using the linear relationship of RH and water content from Table \ref{table:rhwatercontent}. The n/a* for basalt, quartz sand, beach sand, and chromite indicates no water was detected for those materials. The n/a$\dagger$ for walnut shells of all sizes, iced tea, and instant coffee indicated other chemical processes take place instead of the water loss process to high temperature, thus we cannot measure the internal water. For iced tea powder, neither the surface nor the internal water can be measured because chemical processes happen at lower temperature. The n/a$\ddagger$ indicates the estimated water content of quartz sand and chromite at the specific RH are acquired from direct measurements rather than the linear relationships in Table \ref{table:rhwatercontent}.}\label{table:tga}\tabularnewline
 \hline
 Material & Size Range & RH & Estimated Total & Surface & Internal  \\  
Name & $\mu$m & (\%)& Water Content (\%) & Water (\%) & Water (\%) \\  
  \hline\hline
  Basalt & 150--250 & 40 & 0.2 & n/a* & n/a* \\
  \hline
  Quartz Sand & 212--250 & 40 & 0.1 $\ddagger$ & n/a* & n/a*\\
 \hline
 Beach Sand & 500--600 & 40 & 0.1 & n/a* & n/a*\\
  \hline
 Chromite & 212--250 & 40 & 0.1$\ddagger$ & n/a* & n/a* \\
 \hline
 \multirow{6}{*}{Walnut Shells} & 125--150 & 30 &  5.0 & 5.7 & n/a$\dagger$\\ 
 & 150--175 & 40 & 6.6 & 7.3 & n/a$\dagger$\\ 
 & 175--250 & 40 & 6.7 & 7.1 & n/a$\dagger$\\ 
 & 500--600 & 40 & 6.5 & 7.2 & n/a$\dagger$\\
 & 707--833 & 40 & 6.7 & 8.2 & n/a$\dagger$ \\
 & 833--1000 & 40 & 6.7 & 7.9 & n/a$\dagger$ \\
 \hline
 Activated Charcoal & 400--841 & 30 & 6.6 & 4.0 & 1.0 \\
 \hline
 Iced Tea Powder& N/A & 40 & 8.3 & n/a$\dagger$ & n/a$\dagger$ \\ 
 \hline
 Instant Coffee &  N/A & 40 & 8.8 & 6.2 & n/a$\dagger$ \\ 
   \hline
\end{longtable}

To understand the equilibration process for low density materials, we exposed dry walnut shells (150--175 $\mathrm{\mu}$m) to ambient air for different lengths of time, and then measured their surface water content through TGA. The results are shown in Table \ref{table:tgawalnut}. The surface water shows the same value for walnut shells exposed for 2 hours and walnut shells exposed for 4 hours, indicating the surface water equilibrates in 2 hrs or less. The walnut shells exposed for 22 hours have a lower surface water content value, which may have been caused by an RH change during the longer time period.
\begin{table}[h!]
\centering
\caption{The measured equilibration process of walnut shells 150--175 $\mu$m. The four walnut shells samples were baked for 24 hrs in a 120$\degree$C oven and then exposed to air (RH$\sim$40\%) for 0, 2, 4, and 22 hrs. The surface water was then separated by TGA analysis.}
\vspace{0.2cm}
 \begin{tabular}{|c|c|c|c|c|} 
 \hline
 Material & Size Range & RH & Exposed Time & Surface \\  
& $\mu$m & (\%)& (hr) & Water (\%) \\  
  \hline\hline
 \multirow{4}{*}{Walnut Shells} & \multirow{4}{*}{150--175} & 40 & 0 & 0.3 \\ 
 &  & 40 & 2 & 5.3 \\ 
 &  & 40 & 4 & 5.3 \\ 
 &  & 40 & 22 & 4.8 \\
   \hline
\end{tabular}
\vspace{-0.2cm}
\label{table:tgawalnut}
\end{table}

Overall, we found that water in low density materials is dominated by surface water, while interior water occupies less than 20\% of the total water content. Also, surface water is adsorbed first when dry materials are exposed to ambient air. Because only surface water would affect the interparticle cohesion, we believe for low density materials, a change of the water content of the materials would change their interparticle cohesion, and may affect threshold wind speed.

\subsection{The effect of water adsorption on threshold wind speed}
For the `wet' walnut shells, the materials sat and were in equilibrium with ambient air (RH varies between 50--60\%). The water content of the `wet' walnut shells before the TWT run was 8.14\% and after the TWT was 7.20\%, which suggests that the materials were dried by the mixture of air in the TWT due to lower concentration of water vapor (see Section 1). During the TWT run, the RH outside the wind tunnel varied between 50.1\% to 51.8\%, while the RH inside the wind tunnel varied between 16.7\% to 36.5\%.

For the `dry' walnut shells run, after drying and cooling down the sediments, the measured water content was 1.29\%. After the bed was prepared and all the TWT runs finished, the water content of the `dry' walnut shells increased to 1.67\%. During this TWT run, the RH outside the wind tunnel varied between 51.6\% to 53.7\%, while the RH inside the wind tunnel varied between 3.7\% to 11.9\%. Note that the RH inside the wind tunnel for the `wet' scenario is larger than the `dry' scenario. So moisture may come out from the `wet' walnut shells, thus increasing the RH inside the TWT.

Table \ref{table:threshold} shows the threshold freestream wind speed for `wet' and `dry' TWT runs at different pressures. The `wet' thresholds for different pressures are consistently a couple percent larger than the `dry' thresholds. However, the differences are smaller than the standard deviations of the threshold wind speed for both `wet' and `dry' runs.

The reason for the similar `wet' and `dry' thresholds may be due to the similar water adsorption behavior of walnut shells to clay minerals. According to the measurements done by Pirayesh et al. (2012), walnut shells consist of 46.6\% of holocellulose, 49.1\% lignin, and 3.6\% ash. Holocelluose is rich in hydroxyl groups, similar to clay minerals (e.g., kaolinite). The free hydroxyl groups of holocelluose can thus adsorb water through hydrogen bonding, creating an adsorption layer covering the particles (Gwon et al., 2010), while lignin cannot adsorb such a layer of water (Nourbakhsh et al., 2011). Thus walnut shells behave like a mixture of `clay' (hollucellulose) and `quartz sand' (lignin). Note that clay also has long equilibration timescales and high water content similar to walnut shells (Ravi et al., 2006). Thus we would expect that walnut shells behave similarly to a clay/quartz mixture when subjected to water. 

According to Fecan et al., (1999), with clay mixed into quartz sand, threshold does not change when the water content of the materials is below the initiation water content (w'). Threshold will only start to increase with increasing water content when the initiation water content is reached. And with increasing clay content in quartz sand, this initiation water content value is higher. This is caused by the different interparticle cohesion schemes of clay and quartz sand. For quartz sand, the interparticle forces are dominated by capillary forces, which are of similar magnitude to the gravity and wind drag forces. Thus with increasing water content, the threshold will increase accordingly for quartz sand. While for clay minerals, when the water content is lower than the initiation water content value (w\textless w'), the interparticle forces are dominated by adsorption forces due to the molecular bonding between the hydroxyl groups in the clay minerals and water. Adsorption forces are much weaker than capillary forces, thus when the water content increases, even though adsorption forces are increasing, the threshold doesn't change significantly. When the initiation water content is reached (w\textgreater w'), the cohesion forces start to be dominated by capillary forces and the threshold wind speed begins to increase with increasing water content. Simply substituting the 46.6\% holocellulose content as the clay content for walnut shells in Equation \ref{eq:fecan}, we get the initiation water content value, w',
\begin{equation}
w'=0.0014*46.6^2+0.17*46.6=11.0
\end{equation}
When water content of walnut shells is lower than 11.0\%, interparticle forces are dominated by adsorption forces, which is much weaker than capillary forces. It is only when it exceeds 11.0\%, that the interparticle forces are dominated by capillary forces and the threshold begins to increase with increasing water content. This comparison could explain the similar threshold results for the `wet' and `dry' walnut shells with water contents of 1.67\% and 7.20\%, respectively, as both values are lower than the initiation water content value.

\setlength\LTleft{-0.4in}
\setlength\LTright{-0.4in}
\begin{longtable}{|c|c|c|c|c|c|}
\caption{The threshold freestream wind speed for `wet' and `dry' TWT runs at different pressures. The standard deviations were calculated using the procedure in Burr et al. (2015a).}\label{table:threshold}\tabularnewline
\hline
Pressure & `Wet' Threshold   & Standard & `Dry' Threshold  & Standard & Difference\\
(bar) & Freestream & Deviation & Freestream & Deviation & (\%) \\
& Wind Speed (m/s) & (\%) & Wind Speed (m/s) & (\%) & \\
\hline
20 & 1.62 & 5.46 & 1.55 & 2.89 & 4.16\\
\hline
15 & 1.76 & 5.08 & 1.71 & 5.15 & 2.89\\
\hline
12.5 & 1.93 & 3.16 & 1.89 & 5.23 & 2.10\\
\hline
8 & 2.37 & 3.55 & 2.36 & 3.42 & 0.38\\
\hline
3 & 3.48 & 3.85 & 3.36 & 1.88 & 3.46\\
\hline
1 & 6.00 & 2.80 & 5.92 & 0.17 &1.36\\
\hline
\end{longtable}

\begin{figure}
\begin{center}
\includegraphics[width=\textwidth]{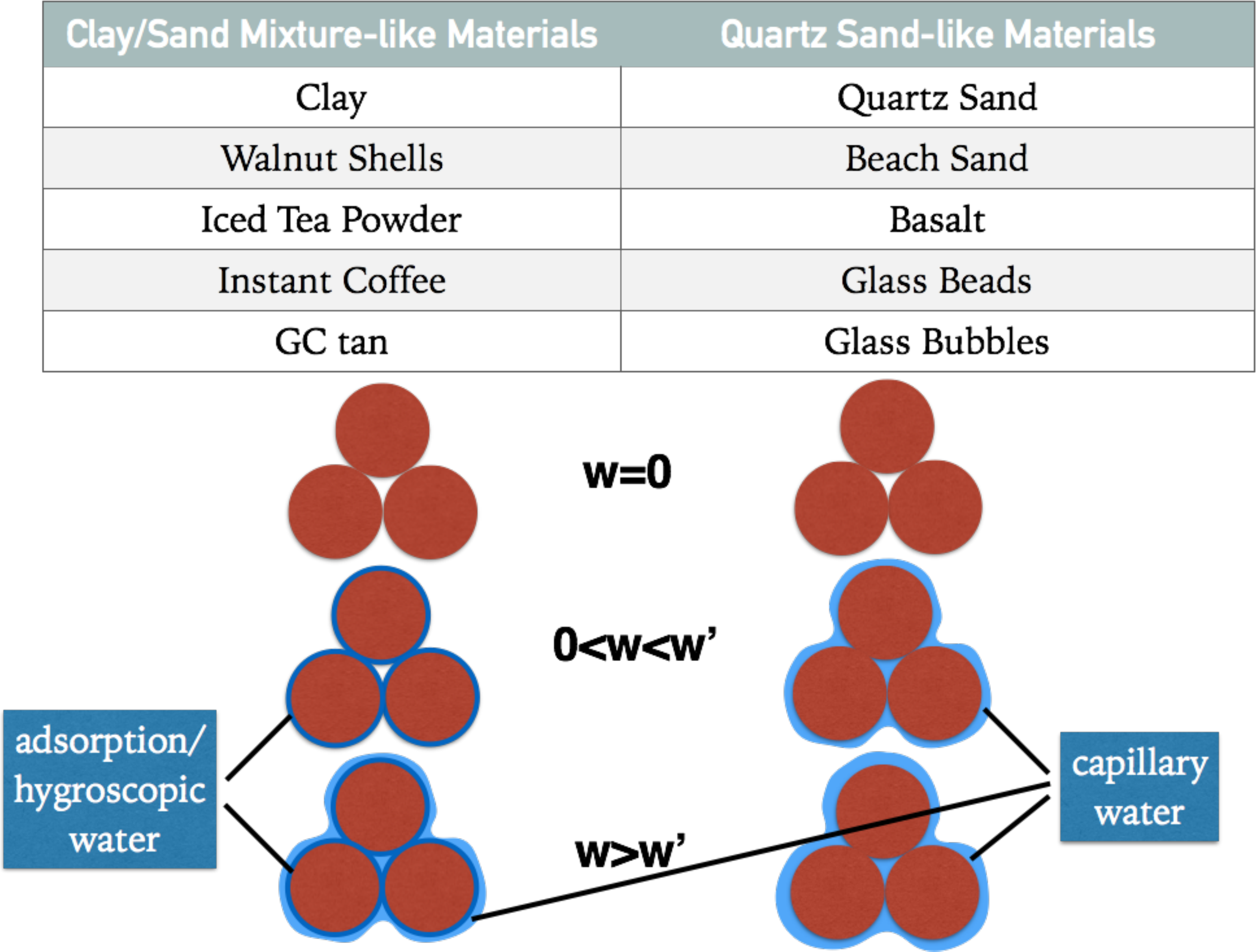}
\end{center}
\caption[]{Lists of clay/quartz sand mixture-like materials and pure quartz sand-like materials and comparison of their behavior when subjected to water. The dark blue layer is the adsorption/hygroscopic water, while the light blue layer is the capillary water. The w is the water content of the material by mass, and w' is the initiation water content.}
\label{fig:watertype}
\end{figure}

The low density materials are usually chosen to match the weight for the relevant planetary body. However, the low density materials may not have the same interparticle forces compared to the real transporting materials. For example, walnut shells used in both the MARSWIT and TWT are more similar to a clay/quartz mixture in terms of interparticle cohesion forces. GC tan (mixture of clay and other minerals), iced tea powder, and instant coffee are similar in that they all adsorb a layer of molecular bonded water (hygroscopic water) and have high water content and low density such that their interparticle forces should behave like walnut shells. They all have an initiation water content, after which the threshold starts to change with increasing water content (see the left column in Fig \ref{fig:watertype}). On Mars, the transporting material is mostly basaltic sand, a high density material, and its interparticle cohesion should be closer to quartz sand. Quartz sand is hydrophobic, so it doesn't form the molecular bonded water layer. Rather, the water on its surface directly contributes to capillary water, as shown on the right column in Fig. \ref{fig:watertype}. Thus with increasing water content of quartz sand, threshold increases accordingly. 

\begin{figure}
\begin{center}
\includegraphics[width=\textwidth]{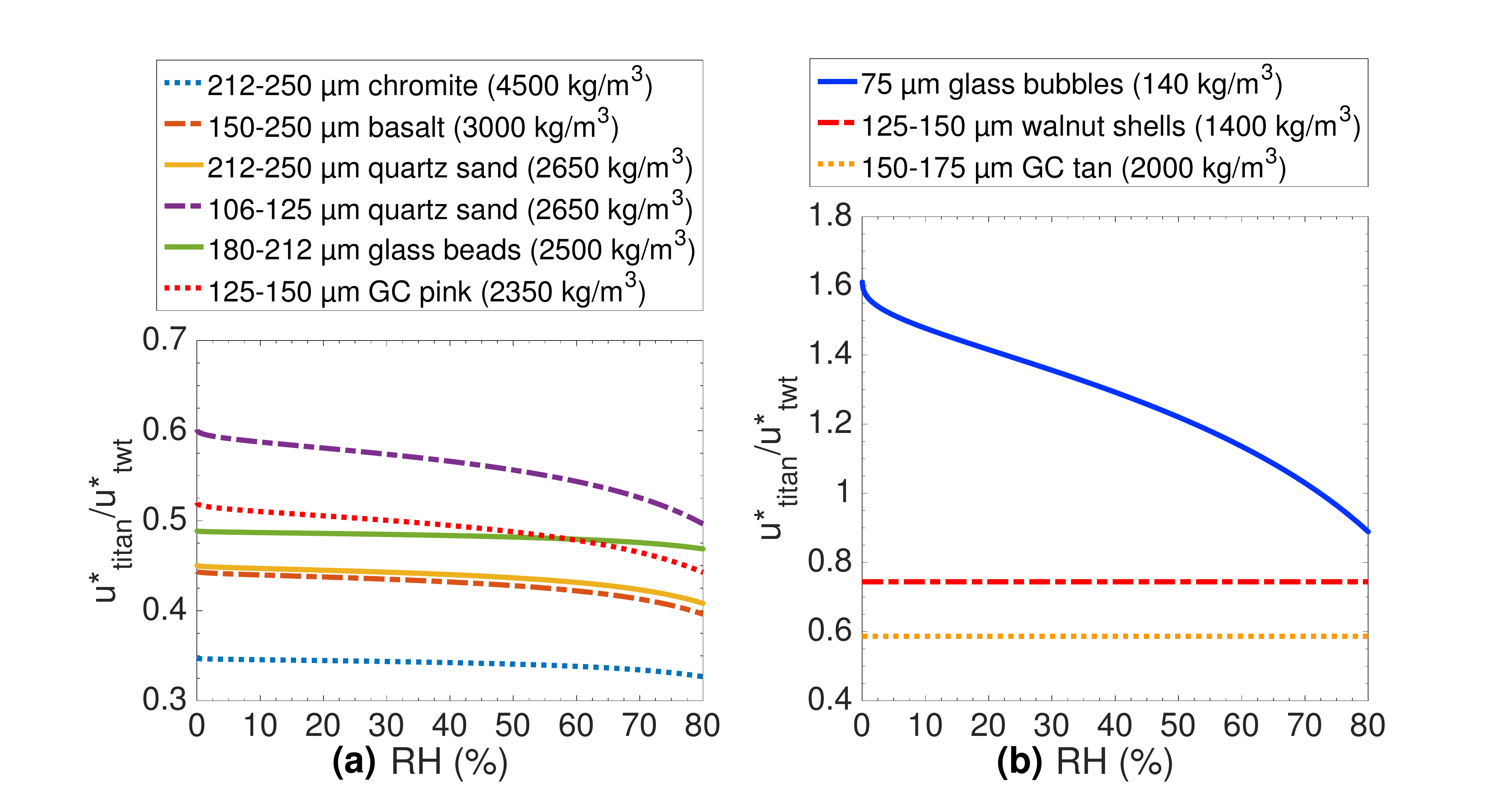}
\end{center}
\caption[]{Modeled threshold wind speed ratios used for converting the $\mathrm{u*_{TWT}}$ to $\mathrm{u*_{Titan}}$ with variation of RH in the TWT between 0 to 80\%, assuming the methane humidity on Titan is 0. (a) For high density materials of different sizes. (b) For part of low density materials, since iced tea powder and instant coffee have unknown size range, we cannot make the theoretical estimation.}
\label{fig:utitanutwt}
\end{figure}

For materials with similar interparticle forces to quartz sand, we can use the model of McKenna Neuman and Sanderson (2008) to translate the TWT results to Titan conditions for different RH in the TWT. These materials have low water content and short equilibration timescales, including all high density materials and one low density material, glass bubbles (see Section 4.2). The conversion ratios to convert $\mathrm{u*_{TWT}}$ to $\mathrm{u*_{Titan}}$ for these materials are shown in Fig. 10(a) and Fig. 10(b, blue line). Materials that are similar to a clay/quartz sand mixtures need a certain water content to alter the interparticle forces from adsorption forces to capillary forces. For walnut shells, this initiation water content is predicted in Section 4.4, 11.0\%. Using the RH-water content relationship for walnut shells in Table 6, we find that the corresponding initiation RH is about 90\%. Since we have never observed such high RH in the TWT, here we only include the density correction (1400 kg/m\textsuperscript{3} instead of 1100 kg/m\textsuperscript{3}), for the conversion ratios in Fig. \ref{fig:utitanutwt}(b). For GC tan, the density correction is 2000 kg/m\textsuperscript{3} instead of 1300 kg/m\textsuperscript{3}. Since no size range is provided for iced tea powder and instant coffee, we cannot make a prediction for them. Activated charcoal is hydrophobic, its high water content is mainly attributed to its large surface area, and currently we cannot conclude which group it belongs to.

To correctly simulate different interparticle force regimes, determination of the water content of the materials is very important. If the water content of the material is high (\textgreater6\%), as is usual for low density materials, the material likely behaves like clay when exposed to water. If the water content of the material is low (\textless1\%), it may behave as quartz sand when exposed to water. Thus the determination of water content could not only distinguish the density of the materials, but more importantly, this information may provide insight on the effect of water RH on threshold. This information makes the only exception, glass bubbles (a low density material with a low water content), very useful in simulating both the gravity and interparticle forces of real transporting Titan particles. Although electrostatic forces can make these glass bubbles stick together (thus make them hard to sieve) and further experiments should consider this effect before using them.

\subsection{The effect of methane humidity on tholins}
On Titan, the transported material is dark organic sand (Barnes et al., 2008), with possible methane and ethane moisture affecting its cohesion force (Lorenz 2014). Laboratory studies of the adsorption of methane and ethane on tholins show that at saturation, tholins can adsorb only 0.3\% of methane by mass (approximately a monolayer) or a monolayer of ethane (Curtis et al., 2008). As the molecular weights of water (18 g/mol) and methane (16 g/mol) are similar, the methane content of tholins may be close to the water content of quartz (and other high density materials). Thus it is possible that the interparticle cohesion of tholins (subjected to methane moisture) is similar to quartz sand (subjected to water vapor); that is, methane acts as capillary liquid instead of an adsorption liquid. With the increasing relative humidity of methane, or increasing methane content of tholins, the threshold wind speed for tholins or Titan's organic sand will increase accordingly, with no initiation liquid content like clay/quartz mixture or walnut shells (to water vapor).

\begin{table}[h!]
\centering
\caption{Modeling parameters for Titan. Hamaker constant for methane is adopted from Iwamatsu and Horii (1996) and Israelachvili (2011).}
\vspace{0.2cm}
 \begin{tabular}{|c|c|} 
 \hline
Modeling Parameter & Value \\  
\hline\hline
Temperature (T) & 94 K \\
\hline
Particle Size ($\mathrm{D_p}$) & 125 $\mathrm{\mu}$m\\
\hline
Air Density ($\mathrm{\rho_a}$) & 5.1 kg/$\mathrm{m^3}$\\
\hline
Particle Diameter ($\mathrm{\rho_p}$) & 950 kg/$\mathrm{m^3}$\\
\hline
Molar Volume of Methane ($\mathrm{V_{CH_4}}$) & 1.6 $\times 10^{-5}$ $\mathrm{m^3/mol}$\\
\hline
Hamaker Constant for Methane (H)* & $\mathrm{-0.5\times10^{-19}}$ J\\
\hline
Roughness Dimensionless Number (k) & $\mathrm{2.1\times10^{-4}}$\\
\hline
Roughness Power (n) & 4.5\\ 
\hline
f(Re*) & 0.024\\
   \hline
\end{tabular}
\vspace{-0.2cm}
\label{table:modelingparameters}
\end{table}

\begin{figure}
\begin{center}
\includegraphics[width=\textwidth]{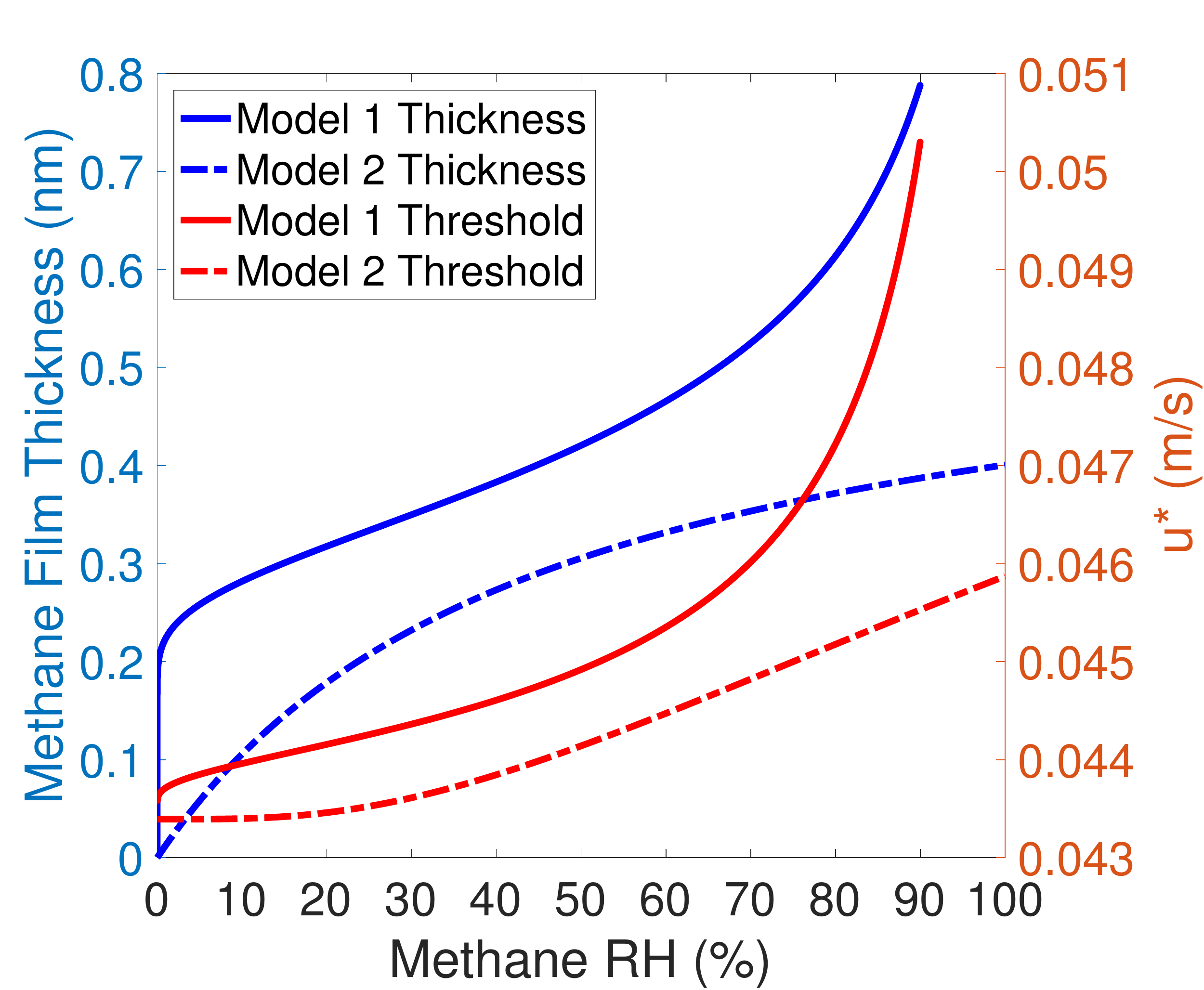}
\end{center}
\caption[]{Two modeling results are shown of methane film thickness for `tholins' and the threshold wind speed variation with changing relative humidity of methane. The blue curves shows the result for the methane film thickness and the red curves are the threshold wind speed variation with methane humidity. Model 1 refers to the revised McKenna Neuman and Sanderson (2008) model. As RH approaches 100\%, the calculated thickness and threshold approach infinity, so here we only show RH between 0 to 90\%. Model 2 shows the thickness and threshold wind speed variation with methane humidity using the data and Langmuir model of Curtis et al. (2008).}
\label{fig:threshold_titan}
\end{figure}

In the definition of the matric potential $\mathrm{\Psi}$ (Equation \ref{eq:psi}), when RH approaches 0, $\mathrm{\Psi \to -\infty}$, and when RH is 100\%, $\mathrm{\Psi=0}$. Thus when including the matric potential in calculating the thickness of the liquid film (Equation \ref{eq:thickness}), the values become extreme when RH is very small or very large (see the solid blue curve in Fig. \ref{fig:threshold_titan}). These extreme RH values would also lead to extreme values for the threshold wind speed u*. To avoid this issue, we developed a second model incorporating measurements of methane film thickness on tholins in Curtis et al. (2008), shown as the dash-dot blue curve in Fig. \ref{fig:threshold_titan}, with a Langmuir adsorption isotherm fit. This thickness fit has no extreme values for the whole RH range and is more realistic compared to McKenna Neuman and Sanderson (2008). To calculate the total interparticle cohesion force, instead of using $\mathrm{\Psi}$ (Equation \ref{eq:fi}) we use another expression of the Laplace pressure $\Delta p$ to avoid the extremes at RH=0 and RH=100\% (Christenson, 1988):
\begin{equation}
\Delta p=\frac{\gamma_s}{r_m}\approx \frac{\gamma_s}{2\delta cos\theta},
\end{equation}
where $\mathrm{\gamma_s}$ is the surface tension of methane, 15 mN/m (Miquet et al., 2000), $\mathrm{\delta}$ is thickness of the methane film, and $\mathrm{\theta}$ is the contact angle between methane and tholins. Here we use $\mathrm{cos\ \theta}=0.97$ (Lavvas et al., 2011).

The results for the computed threshold variation with changing the relative humidity of methane are shown in Fig. \ref{fig:threshold_titan}. The solid red curve shows the modeling result of threshold wind speed variation with RH of methane under Titan conditions using the model of McKenna Neuman and Sanderson (2008), and the dash-dot red curve shows the modeling result incorporating the Langmuir model and data of Curtis et al. (2008). The solid red curve displays a more dramatic change with increasing methane RH than the dash-dot red curve; however, for both models, extreme methane humidity causes the threshold wind speed to change by less than 20\%, compared to dry conditions (RH=0). This minimal change could be attributed to the lower surface tension of methane compared to water and Titan's low temperature. However, this explanation assumes the geometry, contacting mechanics, and electrostatic forces of Titan's organic sand is similar to Earth quartz, which is not known. Thus further research on these properties of tholins is necessary.


\section{Conclusion}
We measured various properties for low density materials used in planetary wind tunnels that have been missing or incomplete in the literature. The literature-given density of walnut shells, in use since the 1970s, 1100 $\mathrm{kg/m^3}$, is lower than our measurement of 1400 $\mathrm{kg/m^3}$, a difference of 30\%. The effect of moisture on low density materials is also very distinct compared to high density materials. Low density materials generally have high water content (\textgreater6\%) and long equilibration timescales (\textgreater6 hrs), while high density materials have low water content (\textless1\%) and short equilibration timescales (\textless1 hr). The determination of the water content of the material provides insight into the sensitivity of threshold wind speed to RH based on our `wet' and `dry' walnut shell TWT runs. The results indicate that threshold is not very sensitive to 'wet' vs 'dry' walnut shells. The materials with high water content tend to behave like a clay/quartz mixture (where adsorption forces dominate below the initiation water content, and then the capillary forces dominate), whereas the materials with low water content are more likely to behave similarly to quartz sand (where capillary forces always dominate). When the interparticle forces are dominated by capillary forces, the threshold increases with increasing water content. Because tholins have a low methane content, we hypothesize that when the real transporting materials on Titan are subjected to methane moisture, they would behave similarly to quartz sand subjected to water.

\section{Acknowledgements}
Partial support to Yu, Bridges, and Burr was provided by NASA grant NNX14AR23G/118460 that was selected under the Outer Planets Research Program (Bridges PI). We thank C. Chavez and J. Moore of NASA Ames Research Center for providing the oven to use in the wind tunnel experiments.



\section{References}
\begin{thebibliography}{00}
\bibitem[Bagnold (1941)]{} Bagnold, R.~A., \ 1941. The physics of wind blown sand and desert dunes. Methuen, London 265(10).
\bibitem[Baidakov et al.(2013)]{} Baidakov, V.~G., Kaverin, A.~M., and Khotienkova, M.~N., \ 2013. Surface tension of ethane methane solutions: 1. Experiment and thermodynamic analysis of the results. Fluid Phase Equilibria, 356, 90. http://dx.doi.org/10.1016/j.fluid.2013.07.008.
\bibitem[Barnes et al.(2008)]{2008Icar..195..400B} Barnes, J.~W., Brown, R.~H., Soderblom, L., et al.\ 2008. Spectroscopy, morphometry, and photoclinometry of Titan's dunefields from Cassini/VIMS. Icarus, 195, 400. http://dx.doi.org/10.1016/j.icarus.2007.12.006.
\bibitem[Belly (1964)]{} Belly, P.~Y., \ 1964. Sand movement by wind. US Army Corps of Engineering (USACE), 1964.
\bibitem[Bird et al.(2005)]{2005Natur.438..800B} Bird, M.~K., Allison, M., Asmar, S.~W., et al., \ 2005. The vertical profile of winds on Titan. Nature, 438, 800. http://dx.doi.org/10.1038/nature04060.
\bibitem[Bisal et al.(1966)]{} Bisal F., and Hsieh, J., \ 1966. Influence of moisture on erodibility of soil by wind. Soil Science, 102(3), 143. 
\bibitem[Bridges et al.(2015)]{2015AGUFM.P12B..05B} Bridges, N., Burr, D.~M., Marshall, J., et al.\ 2015, New Titan Saltation Threshold Experiments: Investigating Current and Past Climates. AGU Fall Meeting Abstracts, P12B-05.
\bibitem[Bunker et al.(2007)]{} Bunker, M.~J., Davies, M.~C., James, M.~B., and Roberts, C.~J, \ 2007. Direct observation of single particle electrostatic charging by atomic force microscopy. Pharmaceutical research, 24(6), 1165. http://dx.doi.org/10.1007/s11095-006-9230-z.
\bibitem[Burr et al.(2015a)]{2015Natur.517...60B} Burr, D.~M., Bridges, N.~T., Marshall, J.~R., et al., \ 2015a. Higher-than-predicted saltation threshold wind speeds on Titan. Nature, 517, 60. http://dx.doi.org/10.1038/nature14088.
\bibitem[Burr et al.(2015b)]{2015aerolianenvironments.517...60B} Burr, D.~M., Bridges, N.~T., Smith, J.~K., et al., \ 2015b. The Titan Wind Tunnel: A new tool for investigating extraterrestrial aeolian environments. Aeolian Research, 18, 205. http://dx.doi.org/10.1016/j.aeolia.2015.07.008.
\bibitem[Cable et al.(2012)]{2012dasdasd} Cable, M.~L., H{\"o}rst, S.~M., Hodyss, R., Beauchamp, P.~M., Smith, M.~A., and Willis, P.~A., \ 2012. Titan tholins: simulating Titan organic chemistry in the Cassini-Huygens era. Chemical Reviews 112(3), 1882. http://dx.doi.org/10.1021/cr200221x.
\bibitem[Charnay et al.(2014)]{2014Icar..241..269C} Charnay, B., Forget, F., Tobie, G., Sotin, C., and Wordsworth, R., \ 2014. Titan's past and future: 3D modeling of a pure nitrogen atmosphere and geological implications. Icarus, 241, 269. http://dx.doi.org/10.1016/j.icarus.2014.07.009.
\bibitem[Chen et al.(1996)]{} Chen, W., Zhibao, D., and Zhenshan, L. et al., \ 1996. Wind tunnel test of the influence of moisture on the erodibility of loessial sandy loam soils by wind. Journal of Arid Environments, 34(4), 391. http://dx.doi.org/10.1006/jare.1996.0119.
\bibitem[Christenson(1988)]{} Christenson, H.~K, \ 1988. Adhesion between surfaces in undersaturated vapors: a reexamination of the influence of meniscus curvature and surface forces. Journal of Colloid and Interface Science, 121(1), 170. http://dx.doi.org/10.1016/0021-9797(88)90420-1.
\bibitem[Clark et al.(2010)]{2010JGRE..11510005C} Clark, R.~N., Curchin, J.~M., Barnes, J.~W., et al.\ 2010. Detection and mapping of hydrocarbon deposits on Titan. Journal of Geophysical Research (Planets), 115, E10005. http://dx.doi.org/10.1029/2009JE003369.
\bibitem[Curtis et al.(2008)]{2008Icar..195..792C} Curtis, D.~B., Hatch, C.~D., Hasenkopf, C.~A., et al., \ 2008. Laboratory studies of methane and ethane adsorption and nucleation onto organic particles: Application to Titan's clouds. Icarus, 195, 792. http://dx.doi.org/10.1016/j.icarus.2008.02.003. 
\bibitem[F{\'e}can et al.(1999)]{1999AnGeo..17..149F} F{\'e}can, F., Marticorena, B., and Bergametti, G., \ 1999. Parametrization of the increase of the aeolian erosion threshold wind friction velocity due to soil moisture for arid and semi-arid areas. Annales Geophysicae, 17, 149. http://dx.doi.org/10.1007/s00585-999-0149-7.
\bibitem[Findor{\'a}k et al.(2016)]{2016...dsad} Findor{\'a}k, R., Fr{\"o}hlichov{\'a}, M., Legemza, J., and Findor{\'a}kov{\'a}, L., \ 2016. Thermal degradation and kinetic study of sawdusts and walnut shells via thermal analysis. Journal of Thermal Analysis and Calorimetry, 1-6. http://dx.doi.org/10.1007/s10973-016-5264-6.
\bibitem[Fulchignoni et al.(2005)]{2005Natur.438..785F} Fulchignoni, M., Ferri, F., Angrilli, F., et al.\ 2005. In situ measurements of the physical characteristics of Titan's environment. Nature, 438, 785. http://dx.doi.org/10.1038/nature04314.
\bibitem[Gladstone et al.(2016)]{2016Sci...351.8866G} Gladstone, G.~R., Stern, S.~A., Ennico, K., et al., \ 2016. The atmosphere of Pluto as observed by New Horizons. Science, 351, aad8866. http://dx.doi.org/10.1126/science.aad8866.
\bibitem[Greeley et al.(1976)]{1976GeophysResLett} Greeley, R., White, B., Leach, R., Iversen, J., and Pollack, J.~B., \ 1976. Mars - Wind friction speeds for particle movement. Geophys. Res. Lett., 3, 417. http://dx.doi.org/10.1029/GL003i008p00417.
\bibitem[Greeley et al.(1977)]{1977dsmc.book.....G} Greeley, R., White, B.~R., Pollack, J.~B., Iverson, J.~D., and Leach, R.~N.,\ 1977. Dust storms on Mars: Considerations and simulations. NASA Tech.~Memo., NASA-TM--78423, 30 p.
\bibitem[Greeley et al.(1980)]{1980GeoRL...7..121G} Greeley, R., Leach, R., White, B., Iversen, J., and Pollack, J.~B., \ 1980. Threshold windspeeds for sand on Mars - Wind tunnel simulations. Geophys. Res. Lett., 7, 121. http://dx.doi.org/10.1029/GL007i002p00121.
\bibitem[Greeley et al.(1984a)]{1984Icar...57..112G} Greeley, R., Iversen, J., Leach, R., et al.\ 1984, Windblown sand on Venus - Preliminary results of laboratory simulations. Icarus, 57, 112.
\bibitem[Greeley et al.(1985b)]{1985pggp.rept..309G} Greeley, R., Marshall, J.~R., and Leach, R.~N.\ 1985, Microdunes and other aeolian bedforms on Venus: Wind tunnel simulations. Reports of Planetary Geology and Geophysics Program.
\bibitem[Greeley and Iversen(1985)]{1985wagp.book.....G} Greeley, R., and Iversen, J.~D.\ 1985. Wind as a geological process on Earth, Mars, Venus and Titan. Cambridge Planetary Science Series, Vol.~4.~Cambridge University Press. 
\bibitem[Gwon et al.(2010)]{2010dada} Gwon, J. ~G., Lee, S.~Y., Chun, S.~J., Doh, G.~H., and Kim, J.~H.,\ 2010. Effects of chemical treatments of hybrid fillers on the physical and thermal properties of wood plastic composites. Composites Part A: Applied Science and Manufacturing, 41(10), 1491. http://dx.doi.org/10.1016/j.compositesa.2010.06.011.
\bibitem[Hirtzig et al.(2013)]{2013Icar..226..470H} Hirtzig, M., B{\'e}zard, B., Lellouch, E., et al.\ 2013. Titan's surface and atmosphere from Cassini/VIMS data with updated methane opacity. Icarus, 226, 470. http://dx.doi.org/10.1016/j.icarus.2013.05.033.
\bibitem[H{\"o}rst and Tolbert(2013)]{2013ApJ...770L..10H} H{\"o}rst, S.~M., and Tolbert, M.~A., \ 2013. In Situ Measurements of the Size and Density of Titan Aerosol Analogs. ApJL, 770, L10. http://dx.doi.org/10.1088/2041-8205/770/1/L10.
\bibitem[Imanaka et al.(2012)]{2012Icar..218..247I} Imanaka, H., Cruikshank, D.~P., Khare, B.~N., and McKay, C.~P., \ 2012. Optical constants of Titan tholins at mid-infrared wavelengths (2.5-25 {$\mu$}m) and the possible chemical nature of Titan's haze particles. Icarus, 218, 247. http://dx.doi.org/10.1016/j.icarus.2011.11.018.
\bibitem[Israelachvili(2011)]{2011sdad} Israelachvili, J.~N., \ 2011. Intermolecular and surface forces. Academic press.
\bibitem[Iversen et al.(1976)]{1976Icar...29..381I} Iversen, J.~D., White, B.~R., Pollack, J.~B., and Greeley, R.\ 1976. Saltation threshold on Mars - The effect of interparticle force, surface roughness, and low atmospheric density. Icarus, 29, 381. http://dx.doi.org/10.1016/0019-1035(76)90140-8.
\bibitem[Iversen and White(1982)]{1982Sedim..29..111I} Iversen, J.~D., and White, B.~R., \ 1982. Saltation threshold on Earth, Mars and Venus. Sedimentology, 29, 111. http://dx.doi.org/10.1111/j.1365-3091.1982.tb01713.x.
\bibitem[Iwamatsu and Horii(1996)]{1996dadsda} Iwamatsu, M., and Horii, K., \ 1996. Capillary condensation and adhesion of two wetter surfaces. Journal of colloid and interface science, 182(2), 400-6. http://dx.doi.org/10.1006/jcis.1996.0480.
\bibitem[Kok et al.(2012)]{2012RPPh...75j6901K} Kok, J.~F., Parteli, E.~J.~R., Michaels, T.~I., and Karam, D.~B., \ 2012. The physics of wind-blown sand and dust. Reports on Progress in Physics, 75, 106901. http://dx.doi.org/10.1111/10.1088/0034-4885/75/10/106901.
\bibitem[Lavvas et al.(2011)]{2011Icar..215..732L} Lavvas, P., Griffith, C.~A., and Yelle, R.~V., \ 2011. Condensation in Titan's atmosphere at the Huygens landing site. Icarus, 215, 732. http://dx.doi.org/10.1016/j.icarus.2011.06.040.
\bibitem[Le Gall et al.(2011)]{2011Icar..213..608L} Le Gall, A., Janssen, M.~A., Wye, L.~C., et al.\ 2011. Cassini SAR, radiometry, scatterometry and altimetry observations of Titan's dune fields. Icarus, 213, 608. http://dx.doi.org/10.1016/j.icarus.2011.03.026.
\bibitem[Lindal et al.(1983)]{1983Icar...53..348L} Lindal, G.~F., Wood, G.~E., Hotz, H.~B., et al., \ 1983. The atmosphere of Titan - an analysis of the Voyager 1 radio occultation measurements. Icarus, 53, 348. http://dx.doi.org/10.1016/0019-1035(83)90155-0.
\bibitem[Lorenz(2014)]{2014Icar..230..162L} Lorenz, R.~D., \ 2014. Physics of saltation and sand transport on Titan: A brief review. Icarus, 230, 162. http://dx.doi.org/10.1016/j.icarus.2013.06.023.
\bibitem[Lorenz et al.(2006)]{2006Sci...312..724L} Lorenz, R.~D., Wall, S., Radebaugh, J., et al., \ 2006. The Sand Seas of Titan: Cassini RADAR Observations of Longitudinal Dunes. Science, 312, 724. http://dx.doi.org/10.1126/science.1123257.
\bibitem[McCord et al.(2006)]{2006P&SS...54.1524M} McCord, T.~B., Hansen, G.~B., Buratti, B.~J., et al., \ 2006. Composition of Titan's surface from Cassini VIMS. Planet. Space Sci., 54, 1524. http://dx.doi.org/10.1016/j.pss.2006.06.007.
\bibitem[McKenna-Neuman and Nickling(1989)]{1989CanadianJournalSedim..29..111I} McKenna-Neuman, C.~M., and Nickling, W.~G., \ 1989. A theoretical and wind tunnel investigation of the effect of capillary water on the entrainment of sediment by wind. Canadian Journal of Soil Science, 69(1), 79. http://dx.doi.org/10.1016/10.4141/cjss89-008.
\bibitem[McKenna Neuman(2003)]{2003Boundarylayer..29..111I} McKenna-Neuman, C.~M., \ 2003. Effects of Temperature and Humidity upon the Entrainment of Sedimentary Particles by Wind. Boundary-Layer Meteorology, 108(1), 61. http://dx.doi.org/10.1023/A:1023035201953.
\bibitem[McKenna Neuman and Sanderson(2008)]{2008JGR..29..111I} McKenna-Neuman, C.~M., and  Sanderson, S., \ 2008. Humidity control of particle emissions in aeolian systems. Journal of Geophysical Research: Earth Surface, 113(F2). http://dx.doi.org/10.1029/2007JF000780.
\bibitem[Miqueu et al.(2000)]{2000dasdad} Miqueu, C., Broseta, D., Satherley, J., Mendiboure, B., Lachaise, J., and  Graciaa, A., \ 2000. An extended scaled equation for the temperature dependence of the surface tension of pure compounds inferred from an analysis of experimental data. Fluid Phase Equilibria, 172(2), 169. http://dx.doi.org/10.1016/S0378-3812(00)00384-8.
\bibitem[Nield et al.(2016)]{2016dsds} Nield, E.~V., Burr, D.~M, Bridges, N.~T, James, J.~K,and  Emery, J.~P, et al., \ 2016. A Wind Tunnel Study of the Effect of Pressure on Saltation Threshold Conditions. LPSC 2016, 47, 1028.
\bibitem[Nourbakhsh et al.(2011)]{2011dsds} Nourbakhsh, A., Baghlani, F.~F, and  Ashori, A., \ 2011. Nano-Si$\mathrm{O_2}$ filled rice husk/polypropylene composites: Physico-mechanical properties. Industrial Crops and Products, 33(1), 183. http://dx.doi.org/10.1016/j.indcrop.2010.10.010.
\bibitem[Pirayesh(2012)]{2012dasd} Pirayesh, H., Khazaeian, A., and  Tabarsa, T., \ 2012. The potential for using walnut (Juglans regia L.) shell as a raw material for wood-based particleboard manufacturing. Composites Part B: Engineering, 43(8): 3276. http://dx.doi.org/10.1016/j.compositesb.2012.02.016.
\bibitem[Radebaugh et al.(2008)]{2008Icar..194..690R} Radebaugh, J., Lorenz, R.~D., Lunine, J.~I., et al., \ 2008. Dunes on Titan observed by Cassini Radar. Icarus, 194, 690. http://dx.doi.org/10.1016/j.icarus.2007.10.015.
\bibitem[Ravi et al.(2004)]{2004GeoRL..31.9501R} Ravi, S., D'Odorico, P., Over, T.~M., and  Zobeck, T.~M., \ 2004. On the effect of air humidity on soil susceptibility to wind erosion: The case of air-dry soils. Geophys. Res. Lett., 31, L09501. http://dx.doi.org/10.1029/2004GL019485.
\bibitem[Ravi et al.(2006)]{} Ravi, S., Zobeck, T.~M., and  Over, T.~M., \ 2006. On the effect of moisture bonding forces in air-dry soils on threshold friction velocity of wind erosion. Sedimentology, 53(3), 597. http://dx.doi.org/10.1111/j.1365-3091.2006.00775.x.
\bibitem[Rodriguez et al.(2014)]{2014Icar..230..168R} Rodriguez, S., Garcia, A., Lucas, A., et al.\ 2014. Global mapping and characterization of Titan's dune fields with Cassini: Correlation between RADAR and VIMS observations. Icarus, 230, 168. http://dx.doi.org/10.1016/j.icarus.2013.11.017.
\bibitem[Sagan and Khare(1979)]{} Sagan, C., and  Khare, B.~N., \ 1979. Tholins - Organic chemistry of interstellar grains and gas. Nature, 277, 102. http://dx.doi.org/10.1038/277102a0.
\bibitem[Selah and Fryrear(1995)]{} Selah, A., and  Fryrear, D.~W., \ 1995. Threshold wind velocities of wet soils as affected by wind blown sand. Soil Science, 160(4), 304. 
\bibitem[Shao and  Lu(2000)]{2000JGR...10522437S} Shao, Y., and  Lu, H., \ 2000. A simple expression for wind erosion threshold friction velocity. J. Geophys. Res., 105, 22437. http://dx.doi.org/10.1029/2000JD900304.
\bibitem[Smith et al.(1989)]{1989Sci...246.1422S} Smith, B.~A., Soderblom, L.~A., Banfield, D., et al., \ 1989. Voyager 2 at Neptune: Imaging Science Results. Science, 246, 1422. http://dx.doi.org/10.1126/science.246.4936.1422.
\bibitem[Soderblom et al.(2007)]{2007P&SS...55.2025S} Soderblom, L.~A., Kirk, R.~L., Lunine, J.~I., et al., \ 2007. Correlations between Cassini VIMS spectra and RADAR SAR images: Implications for Titan's surface composition and the character of the Huygens Probe Landing Site. Planet. Space Sci., 55, 2025. http://dx.doi.org/10.1016/j.pss.2007.04.014.
\bibitem[Stern et al.(2015)]{2015Sci...350.1815S} Stern, S.~A., Bagenal, F., Ennico, K., et al., \ 2015. The Pluto system: Initial results from its exploration by New Horizons. Science, 350, aad1815. http://dx.doi.org/10.1126/science.aad1815.
\bibitem[Thomas et al.(2015)]{2015Sci...347a0440T} Thomas, N., Sierks, H., Barbieri, C., et al., \ 2015. The morphological diversity of comet 67P/Churyumov-Gerasimenko. Science, 347, aaa0440. http://dx.doi.org/10.1126/science.aaa0440.
\bibitem[Tokano(2010)]{2010AeoRe...2..113T} Tokano, T., \ 2010. Relevance of fast westerlies at equinox for the eastward elongation of Titan's dunes. Aeolian Research, 2, 113. http://dx.doi.org/10.1016/j.aeolia.2010.04.003.
\bibitem[Tuller and  Or(2005)]{2005WRR....4109403T} Tuller, M., and  Or, D., \ 2005. Water films and scaling of soil characteristic curves at low water contents. Water Resources Research, 41, W09403. http://dx.doi.org/10.1029/2005WR004142.
\bibitem[Webb(2001)]{2001MircromeriticsInstru} Webb, P.~A., \ 2001. Volume and density determinations for particle technologists. Micromeritics Instrument Corp, 2(16).


\end{thebibliography}


\end{document}